\documentclass{article}

\usepackage{arxiv}

\usepackage[utf8]{inputenc} 
\usepackage[T1]{fontenc}    
\usepackage{hyperref}       
\usepackage{url}            
\usepackage{booktabs}       
\usepackage{amsfonts}       
\usepackage{nicefrac}       
\usepackage{microtype}      
\usepackage{lipsum}
\usepackage{amsthm}

\input{defi}

\title{Assessing Differentially Private Variational Autoencoders under Membership Inference}

\author{
  Daniel Bernau\thanks{Authors contributed equally.}, Jonas Robl\footnotemark[1] \\
  SAP\\
  Karlsruhe, Germany \\
  \texttt{firstname.lastname@sap.com} \\
  \And
  Florian Kerschbaum \\
  University of Waterloo \\
  Waterloo, Canada \\
\texttt{florian.kerschbaum@uwaterloo.ca} \\
}

\begin{document}
\maketitle

\begin{abstract}
We present an approach to quantify and compare the privacy-accuracy trade-off for differentially private Variational Autoencoders. Our work complements previous work in two aspects. First, we evaluate the the strong reconstruction MI attack against Variational Autoencoders under differential privacy. Second, we address the data scientist's challenge of setting privacy parameter \eps, which steers the differential privacy strength and thus also the privacy-accuracy trade-off. In our experimental study we consider image and time series data, and three local and central differential privacy mechanisms. We find that the privacy-accuracy trade-offs strongly depend on the dataset and model architecture. We do rarely observe favorable privacy-accuracy trade-off for Variational Autoencoders, and identify a case where LDP outperforms CDP.
\end{abstract}
\section{Introduction}
Generative machine learning models such as Variational Autoencoders (VAE) and Generative Adversarial Networks (GAN) infer rules about the distribution of training data to generate new images, tables or numeric datasets that follow the training data distribution. The decision whether to use GAN or VAE depends on the learning task and dataset. However, similar to machine learning models for classification~\cite{CLK+18,FJR15,NSH19,SSSS17,ZLH19} trained generative models leak information about individual training data records~\cite{CYZF20,HMDC19,HHB19}. Anonymization of the training data or a training optimizer with differential privacy (DP) can reduce such leakage by limiting the privacy loss that an individual in the training would encounter when contributing their data~\cite{ACM+16,BRG+21,JE19}. Depending on the privacy parameter \eps differential privacy has a significant impact on the accuracy of the generative model since the perturbation affects how closely generated samples follow the training data distribution. Balancing privacy and accuracy for differentially private generative models is a challenging task for data scientist since privacy parameter \eps states an upper bound on the privacy loss. In contrast, quantifying the privacy loss under a concrete attack such as membership inference allows to quantify and compare the accuracy-privacy trade-off between differentially private generative models. 

This paper compares the privacy-accuracy trade-off for differentially private VAE. This is motivated by previous work that has identified VAE are more prone to membership inference attacks than GAN~\cite{HHB19}. Hence, data scientists may want to particularly consider the use of differential privacy when training VAE. In particular, we formulate an experimental study to validate whether our methodology allows to identify sweet spots w.r.t.~the privacy-accuracy trade-off in VAE. We conduct experiments for two datasets covering image and activity data, and for three different local and central differential privacy mechanisms. We make the following contributions:
\begin{itemize}
    \item Quantifying the privacy-accuracy trade-off under membership inference attacks for differentially private VAE.
    \item Comparing local and central differential privacy w.r.t.~the privacy-accuracy trade-off for image and motion data VAE.
\end{itemize}
This paper is structured as follows. Preliminaries are provided in Section~\ref{sec:prel}. We formulate our approach for quantifying and comparing the privacy-accuracy trade-off for DP VAE in Sections~\ref{sec:methodology}. Section~\ref{sec:datasets} introduces reference datasets and learning tasks. Section~\ref{sec:eval} presents the evaluation and is followed by a discussion in Section~\ref{sec:discussion}. We discuss related work in Section~\ref{sec:rel_work}. Section~\ref{sec:conc} provides conclusions.
\section{Preliminaries}
\label{sec:prel}
In the following we provide preliminaries on VAE, MI and DP.

\subsection{Variational Autoencoders}
\label{sec:prel:vae}
Generative models are trained to learn the joint probability distribution $p(X,Y)$ of features $X$ and labels $Y$ of a training dataset $\cali{D}^{train}$. We focus on Variational Autoencoders~(VAE)~\cite{KW14} as generative model. VAE consist of two neural networks: encoder $E$ and decoder $D$. During training a record $x$ is given to the encoder which outputs the mean $E_\mu(x)$ and variance $E_\Sigma(x)$ of a Gaussian distribution. A latent variable $z$ is then sampled from the Gaussian distribution $N(E_\mu(x),E_\Sigma(x))$ and fed into the decoder $D$. After successful training the reconstruction $D(z)$ should be close to $x$. During training two terms are minimized. First, the \textit{reconstruction error} $\lVert D(z)-x \rVert.$ Second, the \textit{Kullback-Leibler divergence} $\textit{KL}(N(E_\mu(x),E_\Sigma(x)) || N(0,1))$ between the distribution of latent variables $z$ and the unit Gaussian. The KL divergence term prevents the network from only memorizing certain latent variables since the distribution should be similar to the unit Gaussian. Kingma et al.~\cite{KW14} motivate the training objective as a lower bound on the log-likelihood and suggest training $E$ and $D$ for a training objective by using the \textit{reparameterization trick}. Samples $D(z)$ are generated from the VAE by sampling a latent variable $z\sim N(0,1)$ and passing $z$ through $D$. Similar to GAN conditional VAE generate samples for a specific label by utilizing a condition $c$ as input to $E$ and $D$.

\subsection{Reconstruction Membership Inference Attack against VAE}
\label{subsec:mi_gen:rec_attack}
Membership inference (MI) attacks against machine learning models aim to identify the membership or non-membership of an individual record w.r.t.~the training dataset $\cali{D}^{train}$ of a target model. To exploit differences in the generated samples of a trained target model the MI adversary $\Ami$ uses a statistical attack model. Therefore, $\Ami$ computes a similarity or error metric for individual records $x$. After having calculated such a metric for a set of records $\Ami$ labels the records with the highest similarity, or lowest error, as members and all other records as non-members. For VAE the reconstruction loss quantifies how close a reconstructed training record is to the original training data record. Based on the reconstruction loss Hilprecht et al.~formulate the reconstruction MI attack against VAE that outperforms prior work~\cite{HHB19}. The reconstruction MI attack assumes that a reconstructed training record will have a smaller reconstruction loss than a reconstructed test record and repeatedly computes the reconstruction $\hat x=D(z)$ for a record $x$ by drawing the latent variable $z$ from the record-specific latent distribution $\cali{N}(\mathbb{E}_\mu(x), \mathbb{E}_\sigma(x))$. The mean reconstruction distance for $N=300$ samples is then calculated by Eq.~\eqref{eq:background:reconstruction}. Furthermore, the reconstruction MI attack depends on the availability of a distance measure $d$. In this work we use the generic Mean Squared Error (MSE) and the image domain specific Structural Similarity Index Measure (SSIM) as distance measures. A record $x$ is likely a training record in case of small mean reconstruction distances for MSE or a similarity close to $1$ for SSIM. 
\begin{align}
    \label{eq:background:reconstruction}
    f_{reconstruction} = - \frac{1}{N} \sum^N_i d(\hat x - x)
\end{align}

\subsection{Differential Privacy}
\label{chap:prel:ldp}
For a dataset \cali{D} differential privacy (DP)~\cite{dwork2006a} can either be used centrally to perturb a function $f(\cali{D})$ or locally to perturb records $x\in\cali{D}$ by perturbation. In central DP (CDP) an aggregation function $f(\cdot)$ is first evaluated and then perturbed by a trusted server. Due to perturbation it is no longer possible for an adversary to confidently determine whether $f(\cdot)$ was evaluated on \cali{D}, or some neighboring dataset \cali{D'} differing in one record. Privacy is provided to records in \cali{D} as their impact on $f(\cdot)$ is limited. Mechanisms \cali{M} that follow Definition~\ref{def:differential-privacy} are used for perturbation of $f(\cdot)$~\cite{dwork2006b}. CDP holds for all possible differences $\|f(\cali{D}) - f(\cali{D'})\|_2$ by scaling noise to the global sensitivity of Definition~\ref{def:gs}. To apply CDP in VAE we use a differentially private version~\cite{ACM+16} of the Adam~\cite{kingma2015} stochastic gradient optimizer\footnote{We used Tensorflow Privacy: \url{https://github.com/tensorflow/privacy}}. We refer to this CDP optimizer as DP-Adam. DP-Adam represents a differentially private training mechanism $\cali{M}_{nn}$ that updates the weight coefficients $\theta_t$ of a neural network per training step $t \in T$ with $\theta_t \leftarrow \theta_{t-1}-\alpha(\tilde g)$, where $\tilde g~=~\cali{M}_{nn}(\partial loss / \partial \theta_{t-1})$ denotes a Gaussian perturbed gradient and $\alpha$ is some scaling function on $\tilde g$ to compute an update, i.e., learning rate or running moment estimations. Differentially private noise is added by the Gaussian mechanism of Definition~\ref{def:dp:gauss}. After $T$ update steps, $\cali{M}_{nn}$ outputs a differentially private weight matrix $\theta$ which is used by the prediction function $h(\cdot)$ of a neural network. DP-Adam bounds the sensitivity of the computed gradients by specifying a clipping norm \cali{C} based on which the gradients get clipped before perturbation. The iterative weight updates during training result in a composition of $\cali{M}_{nn}$ until training step $T$ at which the final private weights $\theta$ are obtained.
We measure the privacy loss under composition by composing the Gaussian mechanism $\sigma$ under Renyi DP~\cite{Mironov17}. We choose this composition theorem over other composition schemes~\cite{ACM+16,KOV17} due to the tighter bound for heterogeneous mechanism invocations. Similar to related work we set $\dlt=\frac{1}{|\cali{D}|}$ in our experiments~\cite{ACM+16,BRG+21}.

\begin{definition}[\epsdlt-Central Differential Privacy] A~mechanism~\cali{M} gives \epsdlt-central differential privacy if $\cali{D},\cali{D'}\subseteq\cali{DOM}$ differing in at most one element, and all outputs $\cali{S}\subseteq\cali{R}$
	 \begin{equation*}
	 \Pr[\cali{M}(\cali{D}) \in \cali{S}] \leq e^{\eps} \cdot \Pr[\cali{M}(\cali{D'}) \in \cali{S}] + \dlt
	 \end{equation*}
	 \label{def:differential-privacy}
\end{definition}
\begin{definition}[Global $\ell_2$ Sensitivity]
Let \cali{D} and \cali{D'} be neighboring. The global $\ell_2$ sensitivity of a function $f$, denoted by $\sens$, is defined as
	\begin{equation*}
		\sens = max_{\cali{D},\cali{D'}}\|f(\cali{D}) - f(\cali{D'})\|_2.
	\end{equation*}
	\label{def:gs}
\end{definition}
\begin{definition}[Gaussian Mechanism~\cite{dwork2014}] Let $\eps\in(0,1)$ be arbitrary. For $c^2>2ln(\frac{1.25}{\dlt})$, the Gaussian mechanism with parameter $\sgm\ge c\frac{\sens}{\eps}$ gives \epsdlt-CDP, adding noise scaled to $\cali{N}(0,\sgm^2)$.
	\label{def:dp:gauss}
\end{definition}

We refer to the perturbation of records $x\in\cali{D}$ as local DP~(LDP)~\cite{wang2017}. LDP is the standard choice when the server which evaluates a function $f(\cali{D})$ is untrusted. In the experiments within this work we use a local randomizer \cali{LR} to perturb each record $x\in\cali{D}$ independently. Since a record may contain multiple correlated features a \cali{LR} must be applied sequentially to each feature which results in a linearly increasing privacy loss. We adapt the definitions of Kasiviswanathan et al.~\cite{KLN+08} in Definition~\ref{def:localrand} to achieve LDP by using \cali{LR}. A series of \cali{LR} executions per record composes to a local algorithm according to Definition~\ref{def:la}. \eps-local algorithms are \eps-local differentially private~\cite{KLN+08}, where \eps~is a summation of all composed \cali{LR} privacy losses. In this work we will use the \cali{LR} by Fan~\cite{Fan18} for LDP image pixelization. Their \cali{LR} applies the Laplace mechanism of Definition~\ref{def:lap-mech} with scale $\lambda=\frac{255 \cdot m}{b^2 \cdot \eps}$ to each pixel. Parameter $m$ represents the neighborhood in which LDP is provided. Full neighborhood for an image dataset would require that any picture can become any other picture. In general, providing DP or LDP within a large neighborhood will require high privacy parameters \eps values to retain meaningful image structure. Small privacy parameters \eps will result in random black and white images. 

\begin{definition}[Local Differential Privacy]
A local randomizer (mechanism) $\cali{LR}: \cali{DOM}\to\cali{S}$ is \eps-local differentially private, if $\eps\geq 0$ and for all possible inputs $v,v'\in \cali{DOM}$ and all possible outcomes $s\in \cali{S}$ of \cali{LR}
	\begin{equation*}
		\Pr[\cali{LR}(v)=s] \le e^{\eps} \cdot \Pr[\cali{LR}(v')=s]
	\end{equation*}
\label{def:localrand}
\end{definition}
\begin{definition}[Local Algorithm] An algorithm is \eps-local if it accesses the database $\cali{D}$ via $\cali{LR}$ with the following restriction: for all $i\in\{1,\ldots,|\cali{D}|\}$, if $\cali{LR}_1(i),\ldots,\cali{LR}_k(i)$ are the algorithms invocations of $\cali{LR}$ on index $i$, where each $\cali{LR}_j$ is an $\eps_j$-local randomizer, then $\eps_1+\ldots+\eps_k\leq\eps$. \label{def:la}
\end{definition}
 \begin{definition}[Laplace Mechanism~\cite{dwork2014}]
\label{def:lap-mech}
Given a numerical query function $f: DOM \rightarrow \mathbb{R}^k$, the Laplace mechanism with $\lambda=\frac{\sens}{\eps}$ is an \eps-differentially private mechanism, adding noise scaled to $Lap(\lambda,\mu=0)$. 
\end{definition}

We furthermore use a domain independent LDP mechanism specifically for VAE, to which we refer as VAE-LDP. VAE-LDP by Weggenmann et al.~\cite{WRA+22} allows a data scientist to use VAE as LDP mechanism to perturb data. 
This is achieved by limiting the encoders mean and adding noise to the encoders standard deviation before sampling the latent code $z$ during training. After training, the resulting VAE is used to perturb records with $\eps = \frac{\sens \sqrt{2\log(1.25/\dlt)}}{\sigma}$. In this work we limit the resulting mean of the encoder to $[-3, 3]$ by using the tanh activation function. Furthermore, we introduce noise $E$ according to noise bound $\sigma$ by enforcing a lower bound on the standard deviation of $E$. We set the standard deviation to $\max(\sigma,v)$.
\section{Accuracy and Privacy for Variational Autoencoders}
\label{sec:methodology}
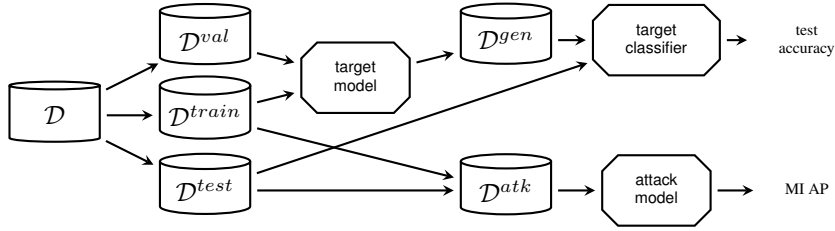
\begin{figure}[ht!]
	\centering
	\begin{tabular}{c}
        \begin{tikzpicture}[thick]
        \tikzstyle{data} = [draw,cylinder,shape border rotate=90,fill=white, shape aspect=.20, minimum width=1.2cm, minimum height=0.8cm, fill=white]
        \tikzstyle{model} = [draw, chamfered rectangle];
        \tikzstyle{arrow} = [->,>=stealth, shorten >=1ex, shorten <=1ex]
        \tikzstyle{larrow} = [->,>=stealth, shorten >=0.5ex, shorten <=0.5ex]
        \tikzstyle{mylabel}=[font=\sffamily]
        
        \node (d) [data] at (0,1) {$\cali{D}$};
        \node (dval) [data] at (2,2) {$\cali{D}^{val}$};
        \node (dtrain) [data] at (2,1){$\cali{D}^{train}$};
        \node (dtest) [data] at (2,0) {$\cali{D}^{test}$};
        \draw[larrow] (d) to (dval);
        \draw[larrow] (d) to (dtrain);
        \draw[larrow] (d) to (dtest);
        \node (target) [model,mylabel] at (4,1.5) {\tiny \begin{tabular}{c} target\\model\end{tabular}};
        \draw[larrow] (dval) to (target);
        \draw[larrow] (dtrain) to (target);
        \node (dgen) [data] at (6,2) {$\cali{D}^{gen}$};
        \draw[larrow] (target) to (dgen);
        \node (classifier) [model,mylabel] at (8,2) {\tiny \begin{tabular}{c} target\\classifier\end{tabular}};
        \draw[larrow] (dgen) to (classifier);
        \draw[larrow] (dtest) to (classifier);
        \node (clres)[] at (10,2){\tiny \begin{tabular}{c} test\\accuracy\end{tabular}};
        \draw[larrow](classifier) to (clres);
        \node (datk) [data] at (6,0) {$\cali{D}^{atk}$};
        \draw[larrow] (dtrain) to (datk);
        \draw[larrow] (dtest) to (datk);
        \node (adv) [model,mylabel] at (8,0) {\tiny \begin{tabular}{c} attack\\model\end{tabular}};
        \draw[larrow] (datk) to (adv);
        \node (atres)[] at (10,0){\tiny \begin{tabular}{c} MI AP\end{tabular}};
        \draw[larrow](adv) to (atres);
        \end{tikzpicture}\\
	\end{tabular}
	\caption{Data flow for the framework.}
	\label{fig:methodology:dataflow}
\end{figure}

We compare the privacy-accuracy trade-off for differentially private VAE to support a data scientist \cali{DS} in choosing privacy parameters \eps. For this we formulate a framework to quantify privacy and accuracy as well as the privacy-accuracy trade-off for differentially private VAE with local or central differential privacy. The framework is depicted in Figure~\ref{fig:methodology:dataflow}. The framework first splits a dataset $\cali{D}$ into three distinct subsets: training data $\cali{D}^{train}$, validation data $\cali{D}^{val}$ and test data $\cali{D}^{test}$. The \textit{target model} VAE is trained on $\cali{D}^{train}$ and optimized on $\cali{D}^{val}$. After training, we use the target model to generate a new dataset $\cali{D}^{gen}$ with the same distribution as $\cali{D}^{train}$. We use $\cali{D}^{gen}$ as input for the \textit{target classifier}, a feed-forward neural network for classification, to quantify the accuracy of the target model by the target classifier accuracy on $\cali{D}^{test}$. Our framework quantifies privacy by means of a MI adversary $\Ami$ performing a MI attack (cf.~Section~\ref{subsec:mi_gen:rec_attack}). The MI attack dataset $\cali{D}^{atk}$ for training and evaluating the MI attack model is sampled equally from $\cali{D}^{train}$ and $\cali{D}^{test}$. We use the framework to calculate the baseline trade-off, as well as CDP and LDP trade-off. The baseline trade-off is calculated from the baseline target classifier test accuracy and the MI attack without any DP mechanism. For the CDP trade-off the target model is trained with DP-Adam (cf.~Section\ref{chap:prel:ldp}).

The LDP trade-off can be computed in three settings to which we refer as LDP-Train, LDP-Full, and VAE-LDP. In LDP-Train a LDP mechanism is applied solely to $\cali{D}^{train}$, but not $\cali{D}^{val}$ and $\cali{D}^{test}$. This scheme is similar to Denoising Autoencoders~\cite{VLL+10}. However, we evaluated the LDP-Train setting and observed it to be mostly impractical for VAE since it introduces a transfer learning task. In particular, working on two different data distributions for $\cali{D}^{train}$ and $\cali{D}^{test}$ leads to distant latent representations and contrasting reconstructions. This neither benefits the target classifier test accuracy nor reduces MI attack performance in comparison to perturbing both training and test data. Hence, we only mention LDP-Train for sake of completeness but will not discuss LDP-Train in the rest of this work. In LDP-Full, $\cali{D}$ is perturbed and the training objective of the target model and the target classifier is changed implicitly (i.e., performance on perturbed data). VAE-LDP perturbs generated data $\cali{D}^{gen}$ by training a perturbation model that follows the target model architecture to enforce LDP. 

The use of LDP also leads to MI attack variations. In particular, the MI attack can either be evaluated against perturbed or unperturbed records in $\cali{D}^{atk}$. We argue that in the LDP-Full setting the MI attack performance against unperturbed records is particularly relevant from the viewpoint of \cali{DS}, since the unperturbed records represent the actual sensitive information and otherwise the attack model would solely learn the to differentiate two distributions by the perturbation skew. Hence, within this work for the LDP settings we exclusively consider the MI attack performance against unperturbed records from $\cali{D}^{train}$.

We evaluate the accuracy of the VAE target model based on the performance of a subsequent target classifier on $\cali{D}^{test}$ after training on $\cali{D}^{gen}$. This is a common approach to evaluate the accuracy of generative models~\cite{FOGD19,JYS19,TKP19}. To evaluate the accuracy of the MI attack we use Average Precision of the Precision-Recall curve (MI AP) which considers membership as sensitive information (i.e., neglecting non-membership). The MI AP quantifies the integral under the precision-recall curve as a a weighted mean of the precision $P$ per threshold $t$ and the increase in recall $R$ from the previous threshold. Using the accuracy of such a curve instead of a singular value allows us to measure the MI attack performance under optimal conditions. For example, the MI adversary $\Ami$ could decide to increase the assumed certainty by raising the threshold closer to 1. Independently of the target model accuracy, \cali{DS} might be interested in lowering MI AP below a predefined threshold that is motivated by legislation (similar to the HIPAA requirement on group sizes~\cite{hhs}).

We quantify the relative trade-off between accuracy and privacy by $\varphi$~\cite{BRG+21} which considers the relative difference between the change in test accuracy for \cali{DS} and the change in MI AP for $\Ami$. We slightly extend the original definition~\cite{BRG+21} to hold for generic accuracy scores that can be used to quantify the accuracy of the target model as well as success of the attacker. Let $ATK$ be a measure to rate the performance of an attack and $ACC$ a measure to rate the performance of the target model. $ATK_{orig}, ACC_{orig}$ represent the scores without DP, while $ATK_{\eps}, ACC_{\eps}$ represent the scores for a specific privacy parameter \eps. Furthermore, let $ATK_{base}, ACC_{base}$ represent the uniform random guessing baseline where $ATK_{base}=0.5$ and $ACC_{base}$ depends on the chosen measure. It applies that $ACC_{base}=\frac{1}{C}$ where $C$ depicts the number of classes. Eq.~\eqref{eq:methodology:phi} provides our adjusted definition for $\varphi$. Similar to the original definition we bound $\varphi$ between $0$ and $2$ s.t.~$\varphi$ does not approach infinity when one measure drops while the other remains stable. $0 \leq \varphi \leq 1$ highlights that the relative loss in model accuracy exceeds the relative loss in attack performance. Contrary, for $1 \leq \varphi \leq 2$ the relative loss in model accuracy is smaller than the relative loss in attack performance. In general, a large gain in privacy, i.e., large drop in attack performance, at a small target model accuracy drop cost is beneficial. Hence \cali{DS} seeks to maximize $\varphi$.
\begin{align}
    \label{eq:methodology:phi}
    \varphi = \min\left(2,\frac{\max(0, (ATK_{orig} - ATK_{\eps})\cdot(ACC_{orig}-ACC_{base}))}{\max(0, (ACC_{orig}-ACC_{\eps})\cdot(ATK_{orig}-ATK_{base}))}\right)
\end{align}
\section{Datasets and Learning Tasks}
\label{sec:datasets}
Within this work we use two reference datasets for image and activity data.

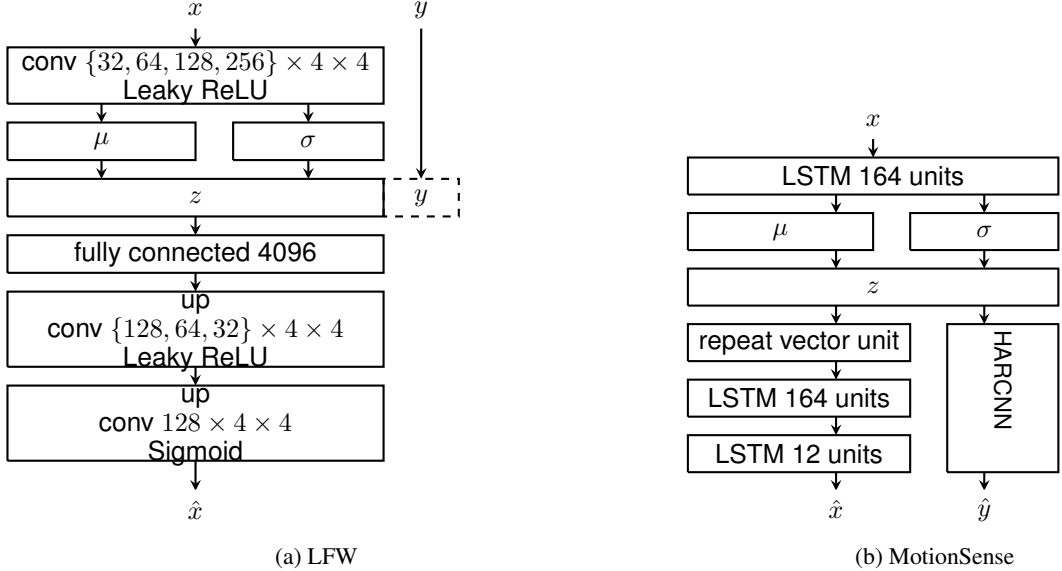
\begin{figure}[ht!]
	\centering
	\begin{subfigure}{0.5\linewidth}
	    \begin{adjustbox}{max width=0.75\textwidth}
        \begin{tikzpicture}[thick]
            \tikzstyle{arrow} = [->,>=stealth]
            \tikzstyle{mylabel}=[font=\sffamily]
            \draw[white] (0,7) -- (0,7.5) -- (5,7.5) -- (5,7) -- (0,7) node [black,midway,yshift=+0.25cm] {$x$};
            \draw[white] (5,7) -- (5,7.5) -- (6,7.5) -- (6,7) -- (5,7) node [black,midway,yshift=+0.25cm] {$y$};
            \draw[arrow] (2.5, 7) -- (2.5,6.75);
            \draw[arrow] (5.5, 7) -- (5.5,5);
            \draw (0,6) -- (0,6.75) -- (5,6.75) -- (5,6) -- (0,6) node [black,midway,yshift=+0.35cm, mylabel,align=center] {conv $\{32,64,128,256\}\times4\times4$\\Leaky ReLU};
            \draw[arrow] (1.25, 6) -- (1.25, 5.75);
            \draw[arrow] (4, 6) -- (4, 5.75);
            \draw (0,5.25) -- (0,5.75) -- (2.5,5.75) -- (2.5,5.25) -- (0,5.25) node [black,midway,yshift=+0.25cm] {$\mu$};
        	\draw[arrow] (1.25, 5.25) -- (1.25, 5);
            \draw (3,5.25) -- (3,5.75) -- (5,5.75) -- (5,5.25) -- (3,5.25) node [black,midway,yshift=+0.25cm] {$\sigma$};
        	\draw[arrow] (4, 5.25) -- (4, 5);
            \draw (0,4.5) -- (0,5) -- (5,5) -- (5,4.5) -- (0,4.5) node [black,midway,yshift=+0.25cm] {$z$};
            \draw[dashed] (5,4.5) -- (5,5) -- (6,5) -- (6,4.5) -- (5,4.5) node [black,midway,yshift=+0.25cm] {$y$};
        	\draw[arrow] (2.5, 4.5) -- (2.5, 4.25);
        	\draw (0,3.75) -- (0,4.25) -- (5,4.25) -- (5,3.75) -- (0,3.75) node [black,midway,yshift=+0.25cm, mylabel] {fully connected 4096};
        	\draw[arrow] (2.5, 3.75) -- (2.5, 3.5);
        	\draw (0,2.5) -- (0,3.5) -- (5,3.5) -- (5,2.5) -- (0,2.5) node [black,midway,yshift=+0.45cm, mylabel,align=center] {up\\conv $\{128,64,32\}\times4\times4$\\Leaky ReLU };
        	\draw[arrow] (2.5, 2.5) -- (2.5, 2.25);
        	\draw (0,1.25) -- (0,2.25) -- (5,2.25) -- (5,1.25) -- (0,1.25) node [black,midway,yshift=+0.45cm, mylabel, align=center] {up\\conv $128\times4\times4$\\Sigmoid};
        	\draw[arrow] (2.5, 1.25) -- (2.5, 0.85);
        	\draw[white] (0,0.35) -- (0, 0.85) -- (5,0.85) -- (5,0.35) -- (0,0.35) node [black,midway,yshift=+0.25cm] {$\hat x$};
        \end{tikzpicture}
        \end{adjustbox}
	\caption{LFW}
	\label{fig:experiments:data:lfw}
	\end{subfigure}%
	\hspace{20 pt}
	\begin{subfigure}{0.4\linewidth}
	  \begin{adjustbox}{max width=0.75\textwidth}
        \begin{tikzpicture}[thick]
            \tikzstyle{arrow} = [->,>=stealth]
            \tikzstyle{mylabel}=[font=\sffamily]
            \draw[white] (0,7) -- (0,9) -- (5,9) -- (5,7) -- (0,7) node [black,midway,yshift=+0.25cm] {$x$};
            \draw[arrow] (2.5, 7) -- (2.5,6.75);
            \draw (0,6.25) -- (0,6.75) -- (5,6.75) -- (5,6.25) -- (0,6.25) node [black,midway,yshift=+0.25cm, mylabel] {LSTM 164 units};
            \draw[arrow] (2, 6.25) -- (2, 6);
            \draw[arrow] (4, 6.25) -- (4, 6);
            \draw (0,5.5) -- (0,6) -- (2.5,6) -- (2.5,5.5) -- (0,5.5) node [black,midway,yshift=+0.25cm] {$\mu$};
        	\draw[arrow] (2, 5.5) -- (2, 5.25);
            \draw (3,5.5) -- (3,6) -- (5,6) -- (5,5.5) -- (3,5.5) node [black,midway,yshift=+0.25cm] {$\sigma$};
        	\draw[arrow] (4, 5.5) -- (4, 5.25);
            \draw (0,4.75) -- (0,5.25) -- (5,5.25) -- (5,4.75) -- (0,4.75) node [black,midway,yshift=+0.25cm] {$z$};
        	\draw[arrow] (2, 4.75) -- (2,4.5);
        	\draw[arrow] (4, 4.75) -- (4, 4.5);
        	\draw (0,4) -- (0,4.5) -- (3,4.5) -- (3,4) -- (0,4) node [black,midway,yshift=+0.25cm, mylabel] {repeat vector unit};
        	\draw[arrow] (2, 4) -- (2, 3.75);
        	\draw (0,3.25) -- (0,3.75) -- (3,3.75) -- (3,3.25) -- (0,3.25) node [black,midway,yshift=+0.25cm, mylabel] {LSTM 164 units};
        	\draw[arrow] (2, 3.25) -- (2, 3);
        	\draw (0,2.5) -- (0,3) -- (3,3) -- (3,2.5) -- (0,2.5) node [black,midway,yshift=+0.25cm, mylabel] {LSTM 12 units};
        	\draw[arrow] (2, 2.5) -- (2, 2.25);
        	\draw[white] (0,1.75) -- (0, 2.25) -- (4,2.25) -- (4,1.75) -- (0,1.75) node [black,midway,yshift=+0.25cm] {$\hat x$};
        	\draw (3.5,2.5) -- (3.5,4.5) -- (5,4.5) -- (5,2.5) -- (3.5,2.5) node [black,midway,yshift=+1.25cm,rotate=270, mylabel] {\small HARCNN};
        	\draw[arrow] (4, 2.5) -- (4, 2.25);
        	\draw[white] (3.5,1.75) -- (3.5,2.25) -- (4.5,2.25) -- (4.5,1.75) -- (3.5,1.75) node [black,midway,yshift=+0.25cm] {$\hat y$};
        \end{tikzpicture}
        \end{adjustbox}
	\caption{MotionSense}
	\label{fig:experiments:data:msvae}
	\end{subfigure}%
	\caption{VAE target model architectures.}
\end{figure}

\paragraph{Labeled Faces in the Wild (LFW).} LFW is a reference dataset for image classification~\cite{HMLL12}. We resize the $250\times250$ images to $64\times64$ by using a bilinear filter and normalize pixels to $[0,1]$ for improved accuracy. Images are distributed unbalanced across the classes with a minimum of 6 and a maximum of 530 pictures. We consider the most frequent 20 and 50 classes to which we refer as LFW20 and LFW50. In total, LFW20 consists of $1,906$ records and LFW50 consists of $2,773$ records. 50\% of the data is allocated to $\cali{D}^{train}$, 20\% to $\cali{D}^{val}$ and 30\% to $\cali{D}^{test}$.
Our VAE target model is an extension of the architecture by Hou et al.~\cite{HSSQ17} and depicted in Figure~\ref{fig:experiments:data:lfw}. $E$ consists of four convolutional layers with $4\times4$ kernels, a stride of two and Leaky ReLU as activation function. $D$ comprises a dense layer followed by four convolutional layers with $3\times3$ kernels, a stride of one and Leaky ReLU as activation function. Before each convolutional layer we perform upsampling by a scale of two with the nearest neighbor method. New data is generated by randomly drawing $z$ from a multivariate Gaussian distribution which is passed through the decoder to create a new record. The target classifier is build upon a pre-trained VGG-Very-Deep-16 (VGG16) model~\cite{SZ15}. The first part of VGG16 consists of multiple blocks of convolutional layers and max-pooling layers for feature extraction. The second part of VGG16 is a fully-connected network for classification. After loading the pre-trained weights\footnote{\url{https://github.com/rcmalli/keras-vggface}} we keep the convolutional core and train the classification part.
\paragraph{MotionSense (MS).} MS is a reference dataset for human activity recognition with $70610$ accelerometer and gyroscope sensor measurements~\cite{MCCH18}. Each measurement consists of twelve datapoints. Measurements are labeled with activities such as walking downstairs, jogging, and sitting. The associated learning task is to label a time series of measurements collected at 50Hz with the corresponding activity. The VAE target model shall reconstruct such a time series. We normalize the data to $[-1,1]$ and group the measurements to time series of 10 seconds. 10\% of the data is allocated to $\cali{D}^{train}$ and $\cali{D}^{val}$ each, and the remaining 80\% is allocated to $\cali{D}^{test}$. Using 10\% of data for training is in line with previous work on MI against generative ML models~\cite{CYZF20,HMDC19,HHB19}. For the target model we use a multitask approach in which $E$ consists of a simple LSTM layer with 164 cells followed by two dense layers for $\mu$ and $\sigma$. $\mu$ and $\sigma$ are used to sample $z$ through the reparameterization trick. $D$ starts with a repeat vector unit for $z$. This allows us to create sequences and pass $z$ to an LSTM layer. Furthermore, a second LSTM layer with twelve units is used to output sequences for each sensor. To support the reconstruction task we input $\mu$ to a classifier. Figure~\ref{fig:experiments:data:msvae} shows the target model architecture. New data is generated by passing training records of a given class $E$ to create $z$, which is then passed through the decoder to generate a record. We have to sample $z$ from the class-specific latent distribution since the latent space is clustered as a consequence of the multitask classifier. The overall loss is balanced with $\lambda_1=0.01$, $\lambda_2=50$, $\lambda_3=0.5$ for KL-loss, reconstruction loss and classifier loss respectively. The target classifier is based on the Human Activity Recognition Convolutional Neural Network (HARCNN) architecture for time series data by Saeed~\cite{saeed2016}. In HARCNN each convolutional layer is followed by a dropout layer which we set to $0.3$ to learn a more general representation of the data. The final two fully-connected layers are used for classification. 
\section{Evaluation}
\label{sec:eval}
Instead of comparing privacy parameter \eps we designed and performed an experiment to compare the privacy-accuracy trade-off in different DP settings. The experiment quantifies the target classifier test accuracy and MI AP by using the framework depicted in Figure~\ref{fig:methodology:dataflow} (cf.~Section~\ref{sec:methodology}). We discuss the experiment for each dataset in four parts. First, we state the \textit{baseline} test accuracy of the target classifier on non-generated data to provide information on the general drop in test accuracy between generated and non-generated data. Second and Third, we discuss CDP and LDP results. Fourth, the results for VAE-LDP are presented. For CDP, LDP and VAE-LDP the experiment results are depicted in two figures each, stating target classifier accuracy over \eps and MI AP over \eps. In each figure we also state the original target classifier test accuracy and MI AP for unperturbed data.

\subsection{Setup}
\label{sec:eval:setup}
For each dataset the target model is trained for $1000$ epochs after which the target model test loss did not decrease significantly while the target classifier accuracy did not increase anymore. The target classifier is trained on generated samples from the VAE until the target classifier test data loss is stagnating (i.e., early stopping). This experiment design avoids overfitting and increases real-world relevance of our results. For CDP we use DP-Adam which samples noise from a Gaussian distribution (cf.~Definition~\ref{def:dp:gauss}) with scale $\sigma = \text{noise multiplier}~z\times \text{clipping norm}~\cali{C}$. We use the heuristic of Abadi et al.~\cite{ACM+16} and set $\cali{C}$ as the median of norms of the unclipped gradients over the course of 100 training epochs. We evaluate increasing CDP noise regimes for the target model by evaluating noise multipliers $z \in \{0.001, 0.01, 0.1, 0.5, 1\}$. The noise levels cover a wide range from baseline accuracy to naive majority vote. The exact $\epsdlt$ values are presented in Table~\ref{tab:experiments:target_model} in the appendix. Due to the varying LDP mechanisms we state the privacy parameter $\eps_i$ for a single mechanism execution for feature $i$ per dataset in the next sections and summarize in \eps in Table~\ref{tab:experiments:target_model}. VAE-LDP perturbation models are trained with various noise bounds $\sigma \in \{0.1, 1, 10, 100, 1000\}$. Again, the corresponding exact $\epsdlt$ values are presented in Table~\ref{tab:experiments:target_model}. For the MI attack we randomly draw $1000$ records both from $\cali{D}^{train}$ and $\cali{D}^{test}$ for $\cali{D}^{atk}$. The experiments were run on Amazon Web Services Elastic Compute Cloud instances of type \enquote{p2.xlarge}\footnote{\url{https://aws.amazon.com/ec2}} with 64 GiB RAM. This instance type is optimized for GPU computing. We implemented our experiments in Python 3.8 and use TensorFlow Privacy\footnote{ \url{https://github.com/tensorflow/privacy}}. We provide all code on GitHub\footnote{\url{https://github.com/SAP-samples/security-research-vae-dp-mia}}. We identify hyperparameter values for batch size, epochs and learning rate for all target classifiers with Bayesian optimization. 

\subsection{LFW}
\label{sec:eval:lfw}
\begin{figure}[ht!]
	\centering
		\begin{subfigure}{0.4\linewidth}
            \includegraphics[width=\linewidth]{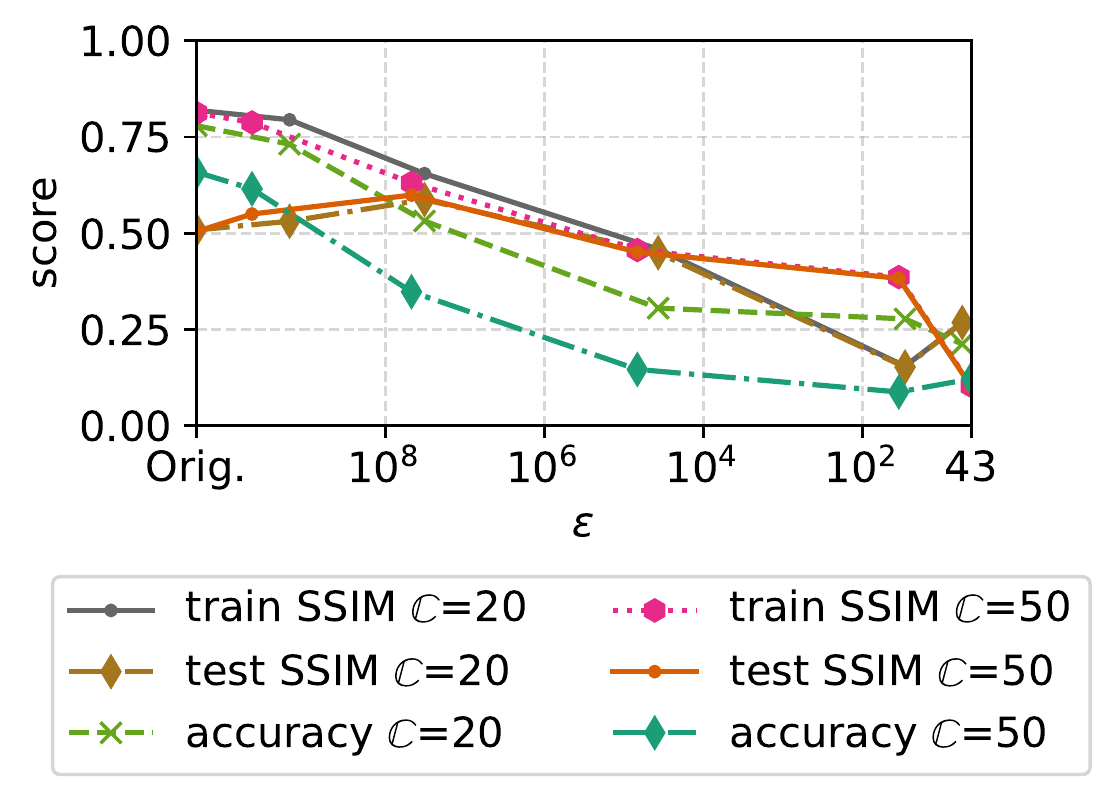}
        	\subcaption{LFW CDP accuracy.}
        	\label{fig:experiments:lfw:cdp:acc}
        \end{subfigure}
    	\begin{subfigure}{0.45\linewidth}
        	\includegraphics[width=\linewidth]{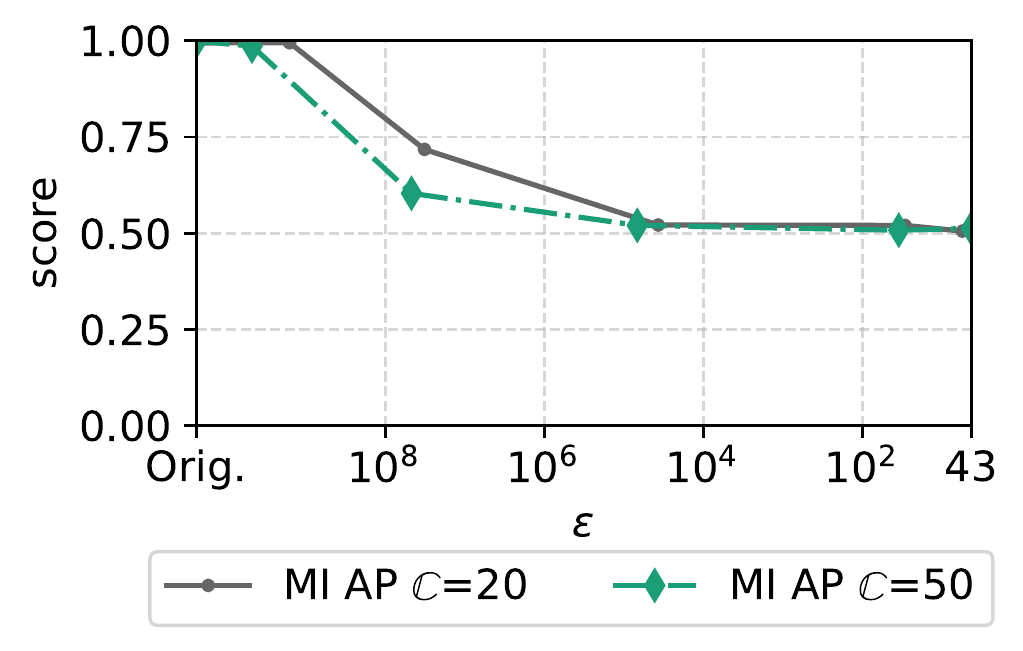}
    	    \subcaption{LFW CDP MI.}
    	    \label{fig:experiments:lfw:cdp:mi}
    	\end{subfigure}\\
		\begin{subfigure}{0.4\linewidth}
            \includegraphics[width=\linewidth]{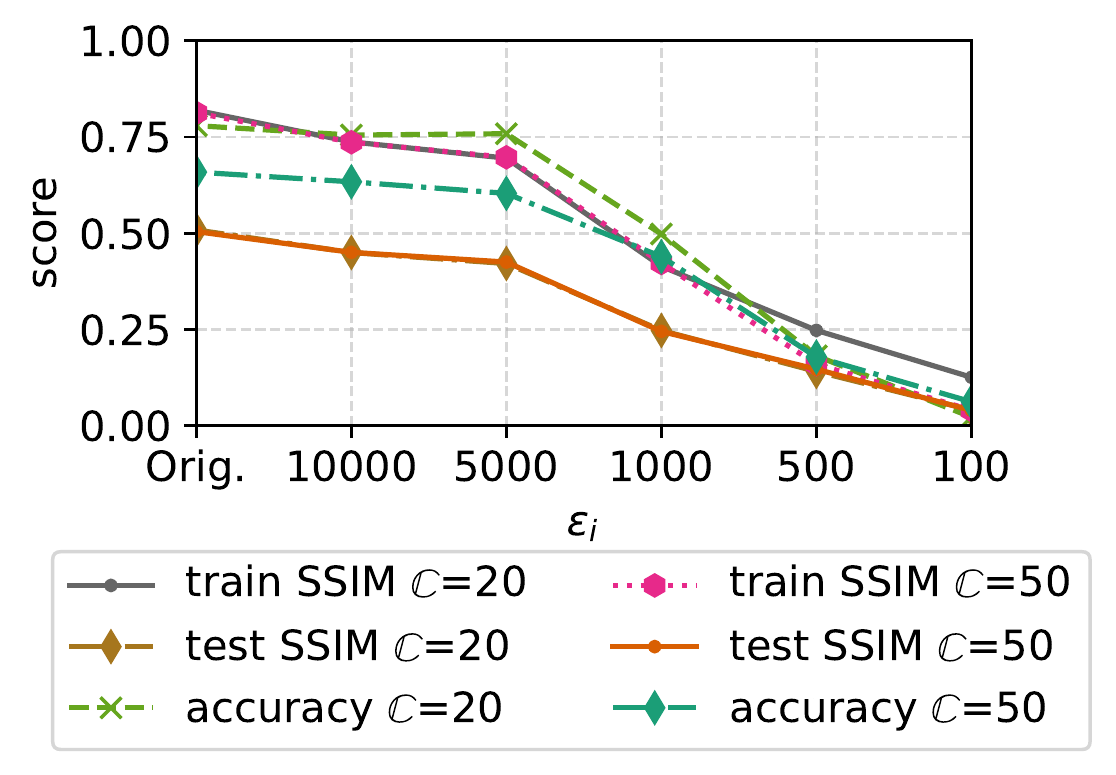}
        	\subcaption{LFW LDP accuracy.}
        	\label{fig:experiments:lfw:ldpfull:acc}
        \end{subfigure}
    	\begin{subfigure}{0.45\linewidth}
        	\includegraphics[width=\linewidth]{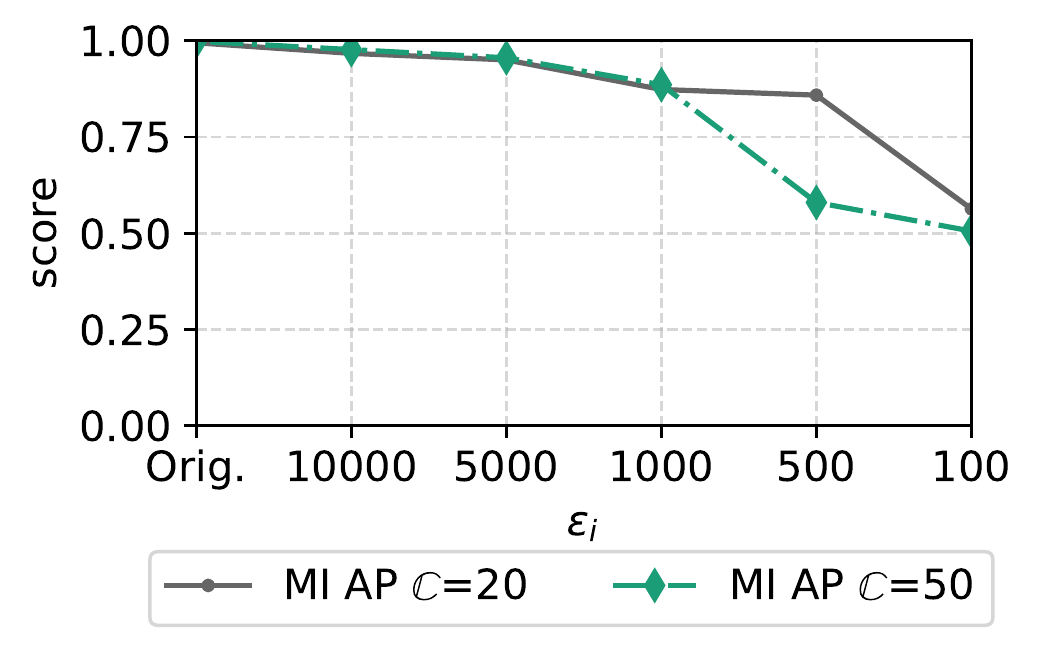}
    	    \subcaption{LFW LDP MI.}
    	    \label{fig:experiments:lfw:ldpfull:mi}
    	\end{subfigure}\\
		\begin{subfigure}{0.4\linewidth}
            \includegraphics[width=\linewidth]{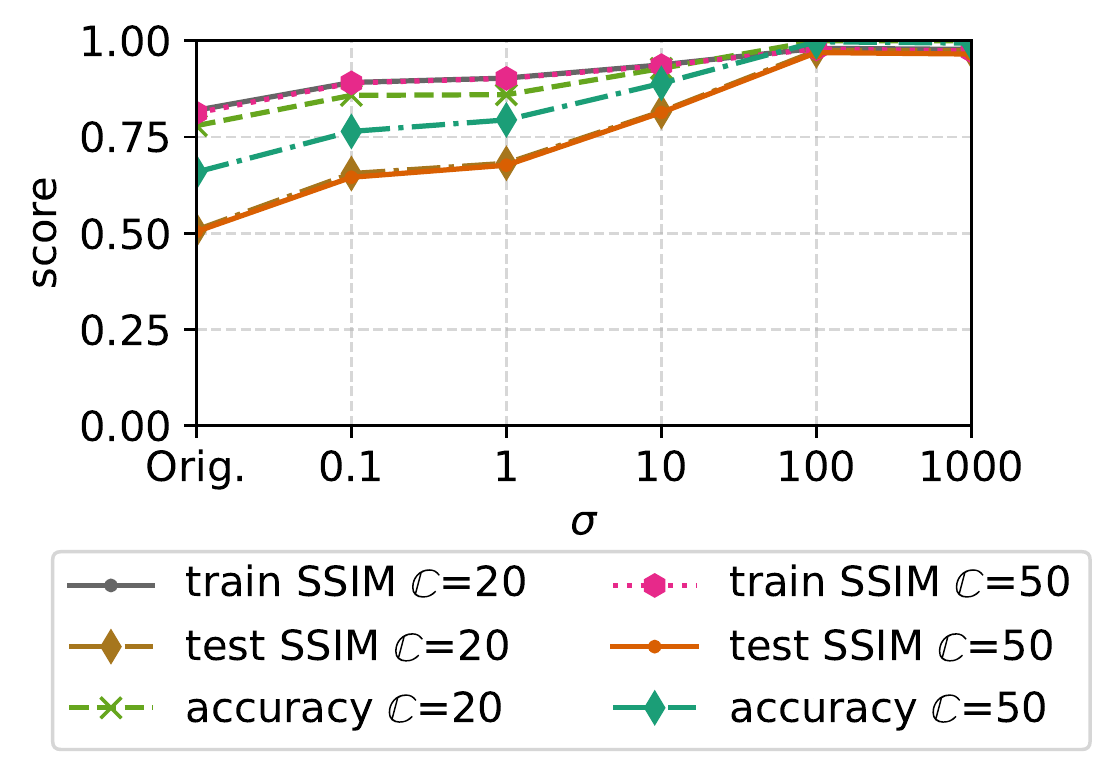}
        	\subcaption{LFW VAE-LDP accuracy.}
        	\label{fig:experiments:lfw:vaeldpfull:acc}
        \end{subfigure}
    	\begin{subfigure}{0.45\linewidth}
        	\includegraphics[width=\linewidth]{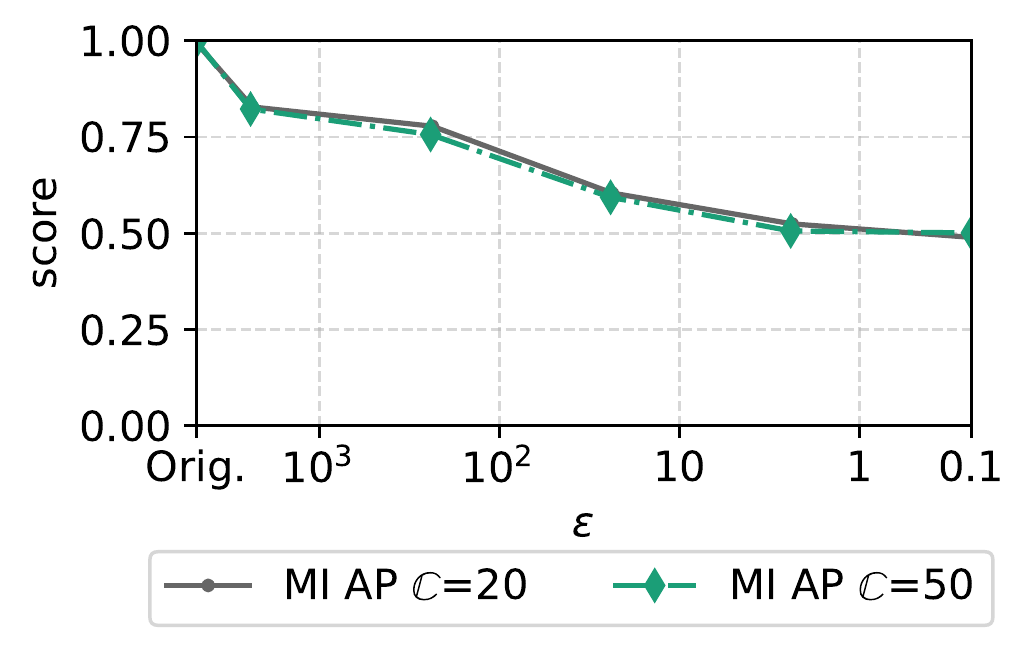}
    	    \subcaption{LFW VAE-LDP MI.}
    	    \label{fig:experiments:lfw:vaeldpfull:mi}
    	\end{subfigure}
    	\caption{LFW accuracy and privacy.}
    	\label{fig:experiments:lfw}
\end{figure}

On non-generated baseline images the target classifier achieves baseline test accuracies of $0.78$ and $0.66$ for LFW20 and LFW50. For generated images we provide two accuracy metrics. Namely, the SSIM of the images generated by the target model and the test accuracy of the target classifier. Figure~\ref{fig:experiments:ms:cdp:acc} states the accuracy metrics for unperturbed and CDP perturbed VAE. The figure illustrates that the unperturbed VAE does not generate images with close proximity to the baseline images. However, the images still suffice to produce target classifier test accuracies well above majority voting. Shapes of the head, hair, and some facial expressions as well as the background can be observed for reconstructed images in Figure~\ref{fig:experiments:samples} in the appendix. We also use SSIM as a domain specific distance metric for the reconstruction MI attack. Figure~\ref{fig:experiments:lfw:cdp:mi} illustrates that the reconstruction MI attack yields a perfect MI AP of 1 for unperturbed VAE. This high MI AP is due to the large gap between train and test SSIM.

Figure~\ref{fig:experiments:ms:cdp:acc} states CDP test accuracy over \eps. The steady accuracy decrease is due to the closing target model train-test gap, which we state in Table~\ref{tab:experiments:target_model} in the appendix. The resulting regularization also lowers the SSIM of the generated images. A particular sharp drop in SSIM is observable for $z=0.5$ ($\eps\approx350$). For this datapoint posterior collapse occurs when $E$ produces noisy $\mu$ and $\sigma$ leading to unstable latent codes $z$ which in turn are ignored by $D$. In consequence, $D$ produces reconstructions independently of $z$ leading to a increased reconstruction loss, while $\mu$ and $\sigma$ become constant and minimize the KL-loss~\cite{LTGN19}. As a consequence the target classifier resorts to majority vote. The CDP MI AP over \eps is stated in Figure~\ref{fig:experiments:lfw:cdp:mi}. The increased regularization caused by CDP is at the same time lowering MI AP. In addition, due to the inherent label imbalance in LFW the VAE reconstruction of loosely populated classes is worse than the reconstruction for classes with more records. Still, the resulting privacy-accuracy trade-off leaves space for compromise. When \cali{DS} would for example be willing to accept an MI AP of up to $0.6$ this would require setting $z\leq0.1$ ($\eps\approx10^5$). $z=0.1$ leads to target classifier test accuracy of $0.31$. However, if \cali{DS} raise their threshold to $0.75$ this would allow for $z=0.01$ ($\eps\approx10^8$) and a target classifier test accuracy of $0.52$. 

For LDP we use differentially private image pixelization (cf.~Section~\ref{chap:prel:ldp}) to create LDP training and test datasets with within neighborhood $m=\sqrt{64\times64}$. Figure~\ref{fig:experiments:lfw:ldpfull:acc} presents the LDP test accuracy and SSIM over $\eps_i$. In contrast to the CDP experiments the target classifier test accuracy and target model SSIM metrics do not show a regularization effect caused by the introduced noise for LDP. The train-test gap narrows only slightly and the random noise introduced in the dataset makes the reconstruction task for the VAE more difficult.  Thus, the reconstruction MI attack AP in Figure~\ref{fig:experiments:lfw:ldpfull:mi} remains nearly unchanged until $\eps_i\le=500$ at which point the target model SSIM and the target classifier test accuracy are already at poor levels and little room for compromise is existing.

VAE-LDP accuracy over \eps is presented in Figure~\ref{fig:experiments:lfw:vaeldpfull:acc}. Counterintuitively, the test accuracy even rises over \eps and the train-test gap and SSIM gap narrow. This is due to the VAE-LDP perturbation model which reconstructs only essential facial features and leaves the background grey when faced with small $\eps$. Hence the learning task for the target classifier the reconstruction task for the VAE are simplified.  Figure~\ref{fig:experiments:samples} in the appendix underlines this observation by showing the same image for VAE-LDP with increasing noise. The reconstruction attack against VAE-LDP in Figure~\ref{fig:experiments:lfw:vaeldpfull:mi} also decreases as the SSIM gap closes. All in all, the results point towards an advantage of the VAE-LDP mechanism over the LDP image pixelization mechanism. The main disadvantage of the VAE-LDP mechanism over image pixelization is the increased effort to optimize perturbation model hyperparameters.

\subsection{MotionSense}
\label{sec:eval:ms}
\begin{figure}[ht!]
	\centering
		\begin{subfigure}{0.4\linewidth}
            \includegraphics[width=\linewidth]{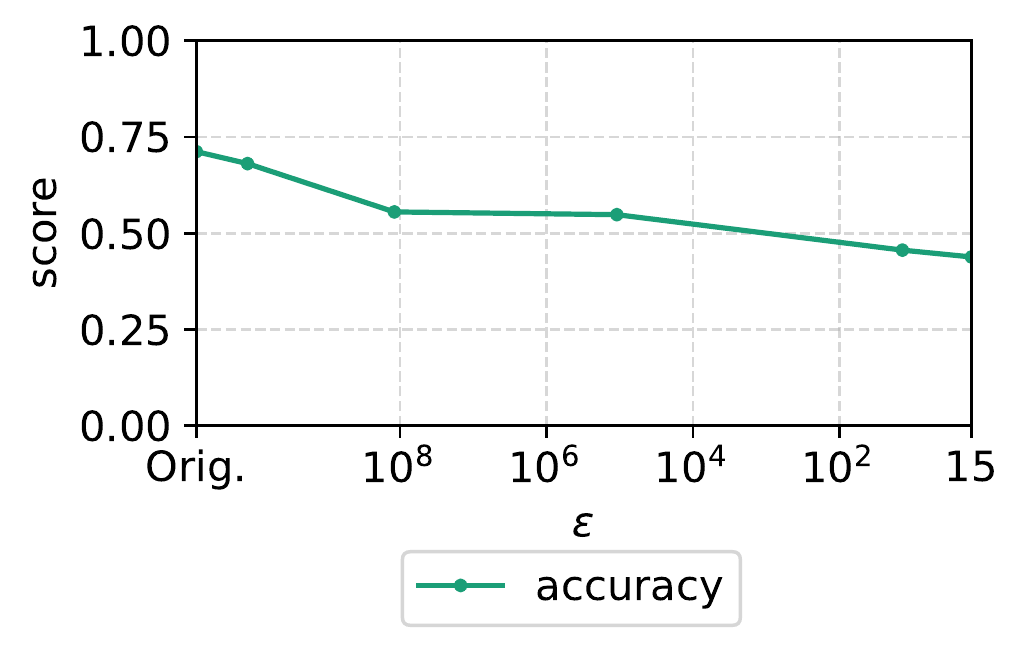} 
        	\subcaption{MS CDP accuracy.}
        	\label{fig:experiments:ms:cdp:acc}
        \end{subfigure}
    	\begin{subfigure}{0.4\linewidth}
        	\includegraphics[width=\linewidth]{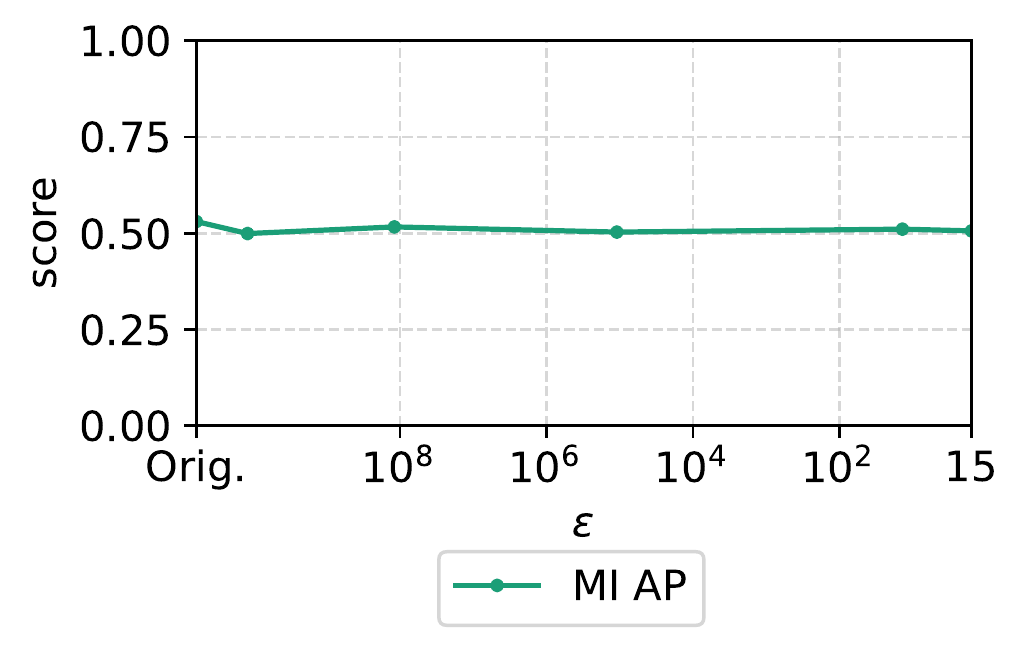}
    	    \subcaption{MS CDP MI.}
    	    \label{fig:experiments:ms:cdp:mi}
    	\end{subfigure}\\
		\begin{subfigure}{0.4\linewidth}
            \includegraphics[width=\linewidth]{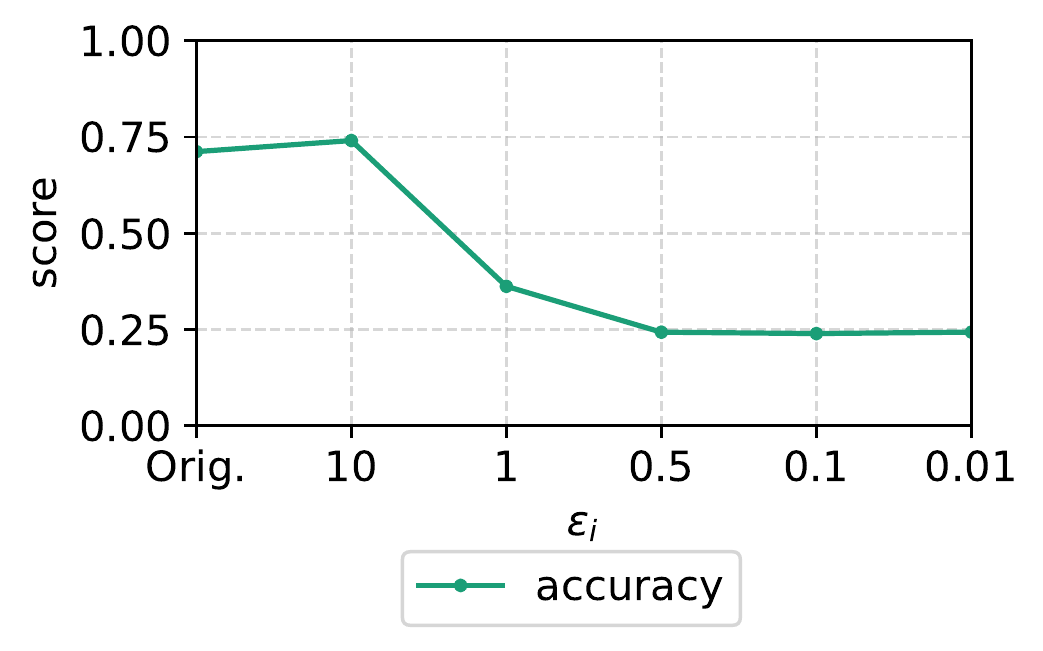} 
        	\subcaption{MS LDP accuracy.}
        	\label{fig:experiments:ms:ldpfull:acc}
        \end{subfigure}
    	\begin{subfigure}{0.4\linewidth}
        	\includegraphics[width=\linewidth]{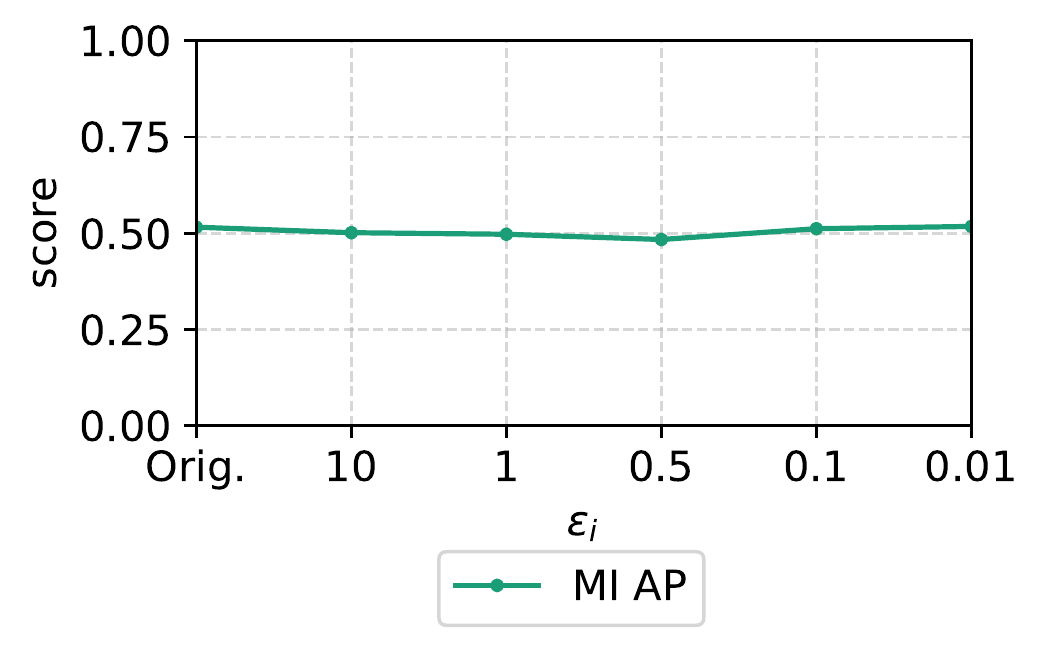}
    	    \subcaption{MS LDP MI.}
    	    \label{fig:experiments:ms:ldpfull:mi}
    	\end{subfigure}\\
		\begin{subfigure}{0.4\linewidth}
            \includegraphics[width=\linewidth]{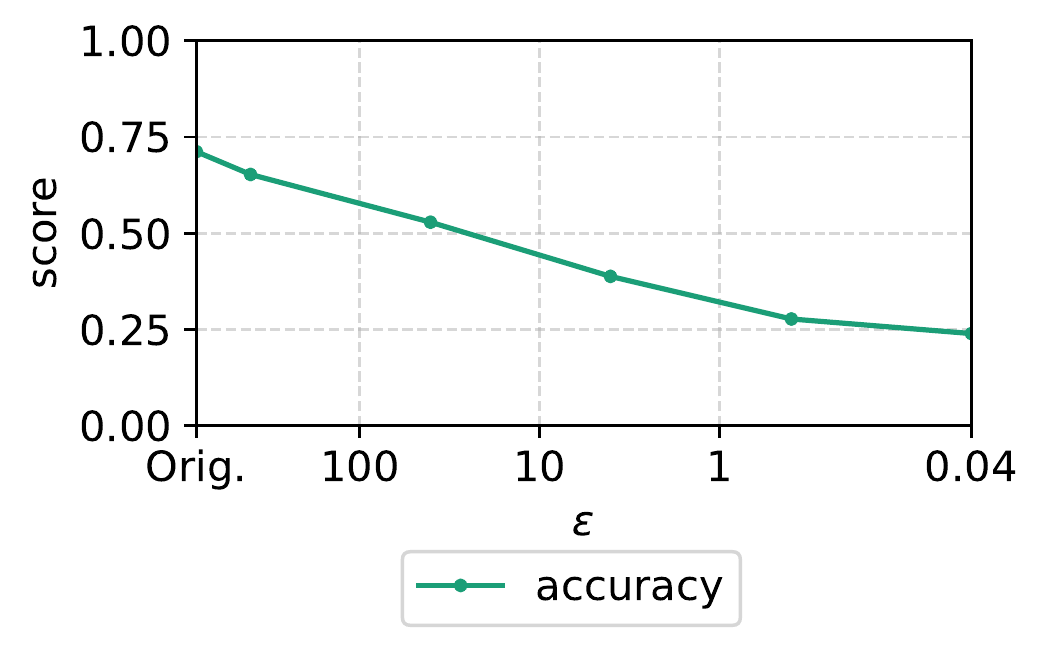}
        	\subcaption{MS VAE-LDP accuracy.}
        	\label{fig:experiments:ms:vaeldpfull:acc}
        \end{subfigure}
    	\begin{subfigure}{0.4\linewidth}
        	\includegraphics[width=\linewidth]{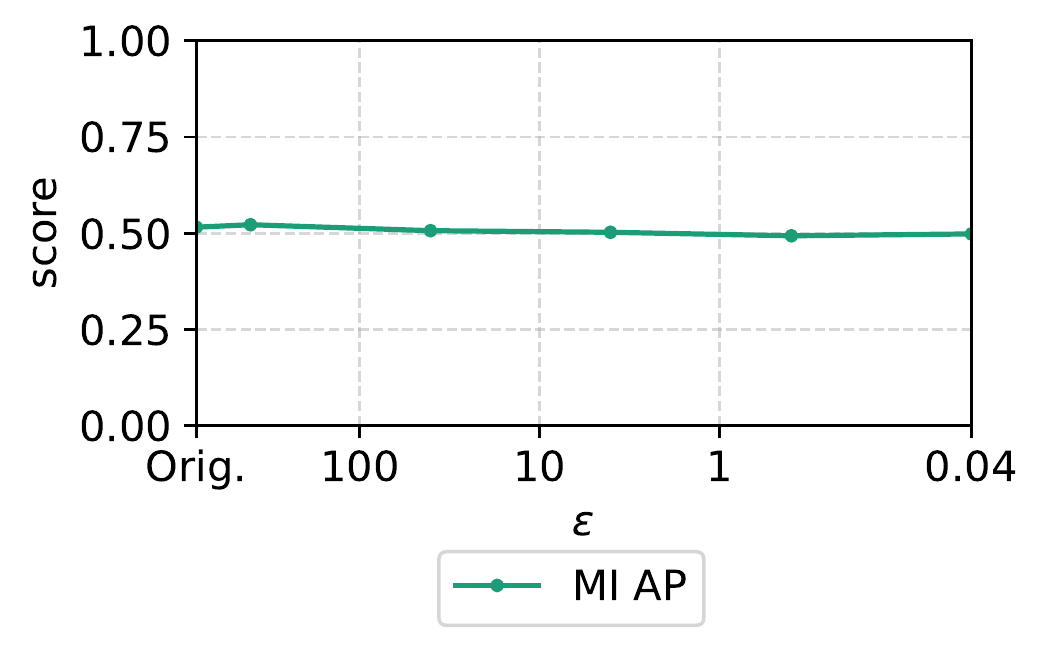}
    	    \subcaption{MS VAE-LDP MI.}
    	    \label{fig:experiments:ms:vaeldpfull:mi}
    	\end{subfigure}
    	\caption{MS accuracy and privacy.}
    	\label{fig:experiments:ms}
\end{figure}

Due to the absence of a domain specific accuracy metric we solely consider test accuracy as accuracy metric for this dataset. The target classifier for MS achieves a baseline test-accuracy of $0.99$ for non-generated data. Figure~\ref{fig:experiments:ms:cdp:acc} states the test accuracy for original and CDP perturbed data over \eps. The test accuracy is dropping to $0.71$ for generated data, which is due to the target model being unable to reconstruct time series for all activities equally well. The reconstruction MI attack has not been used for a time series data in previous work and we suggest ti use MSE as reconstruction MI attack distance metric. The original MI attack performance is depicted in Figure~\ref{fig:experiments:ms:cdp:mi} and achieves an MI AP $0.52$. We see three main reasons for the low MI AP in comparison to LFW. First, MS is more balanced in comparison to LFW. Second, there are significantly more records in MS than in LFW and thus more records per class allow to learn a more general representation. Third, sensor measurements exhibit ambiguities and thus the target model tends to learn more general trends instead of absolute values. 

The CDP target classifier test accuracy only slightly worsens with increasing noise as illustrated in Figure~\ref{fig:experiments:ms:cdp:acc}. This is mostly due to the target classifier resorting to majority vote for particular activities with increasing noise. Figure~\ref{fig:experiments:cdp:ms:cm} in the appendix shows the confusion matrix for the target classifier at $z=1$ ($\eps\approx16$). The target classifier resorts to majority vote for classes 0 to 3 which represent different types of movements, but is still able to distinguish classes $4$ and $5$ which represent standing and sitting. The latter two activities are of different nature than the movements and remain distinguishable under noise. The MI AP illustrated in Figure~\ref{fig:experiments:ms:cdp:mi} shows again the ineffectiveness of the reconstruction MI attack against the MS time series data.

For LDP we use the Laplace mechanism to perturb each measurement (cf.~Section~\ref{chap:prel:ldp}) and specify the sensitivity per sensor as the maximum of all corresponding observed values to create differentially private time series. Figure~\ref{fig:experiments:ms:ldpfull:acc} shows the target classifier accuracy over $\eps_i$. Notably, the target classifier test accuracy increases slightly before dropping sharply over $\eps_i$. Here, small noise levels are actually positively influencing the target model training and hence also allow the target classifier to better distinguish between different classes. In general, the simple LDP mechanism used within this experiment seems to prevents the target model to infer structural information and in turn limits reconstruction or and meaningful generation of records. Figure~\ref{fig:experiments:ms:ldpfull:mi} presents the MI attack performance. The MI AP decreases to $0.5$ already at the largest $\eps_i$ and remains close to the baseline for all further $\eps_i$. 

VAE-LDP test accuracy over \eps is depicted in Figure~\ref{fig:experiments:ms:vaeldpfull:acc}. In comparison to LFW the MS perturbation models do not focus on the essential features of the data and in turn the target classifier cannot benefit from increased perturbation. Due to this the predictions also shift to a majority vote for class $5$ and lower the test accuracy significantly. The VAE-LDP MI AP over \eps is illustrated in Figure~\ref{fig:experiments:ms:vaeldpfull:mi}. Note that at $\sigma=0.1$ ($\eps\approx40$) an outlier is present where the target model did not learn a continuous latent space and thus the reconstruction of records from $\cali{D}^{test}$ suffered. However, the VAE-LDP results show similar trends as the above LDP results.
\section{Discussion}
\label{sec:discussion}
This section discusses the findings of this paper w.r.t.~comparing the privacy-accuracy trade-off for differentially private VAE.
\begin{figure}[ht!]
	\centering
		\begin{subfigure}{0.33\linewidth}
            \includegraphics[width=\linewidth]{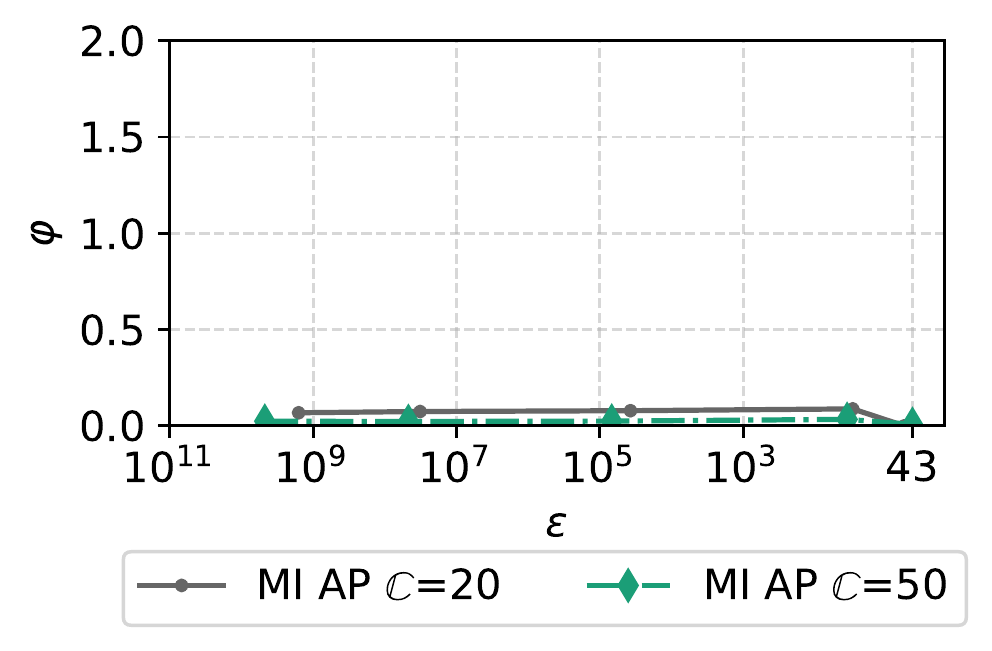}
        	\subcaption{LFW CDP.}
        	\label{fig:discussion:phi:cdp:lfw}
        \end{subfigure}%
        \begin{subfigure}{0.33\linewidth}
        	\includegraphics[width=\linewidth]{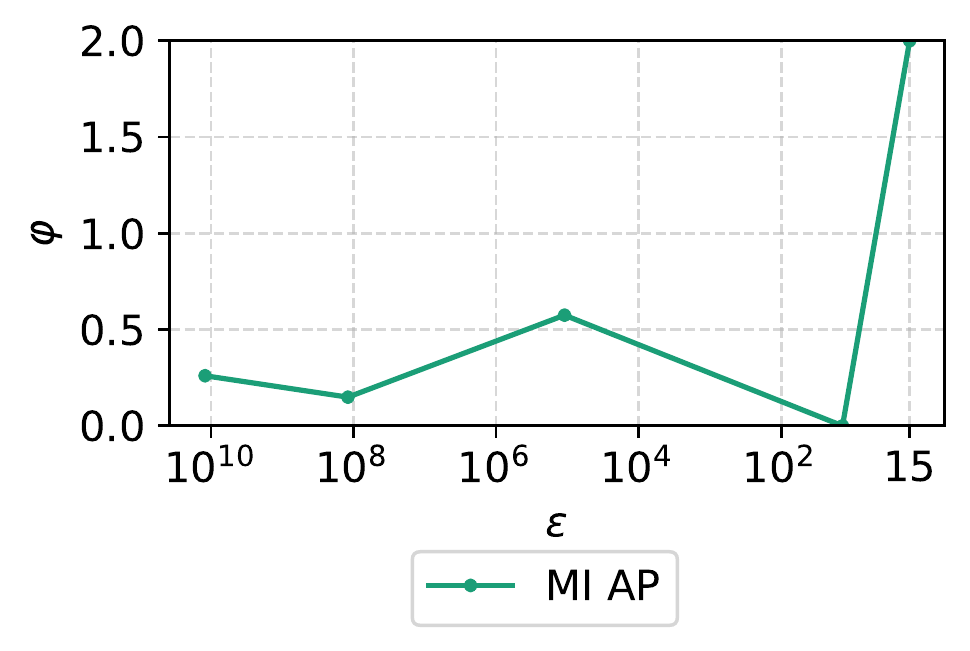}
    	    \subcaption{MS CDP.}
    	    \label{fig:discussion:phi:cdp:ms}
    	\end{subfigure}%
		\begin{subfigure}{0.33\linewidth}
            \includegraphics[width=\linewidth]{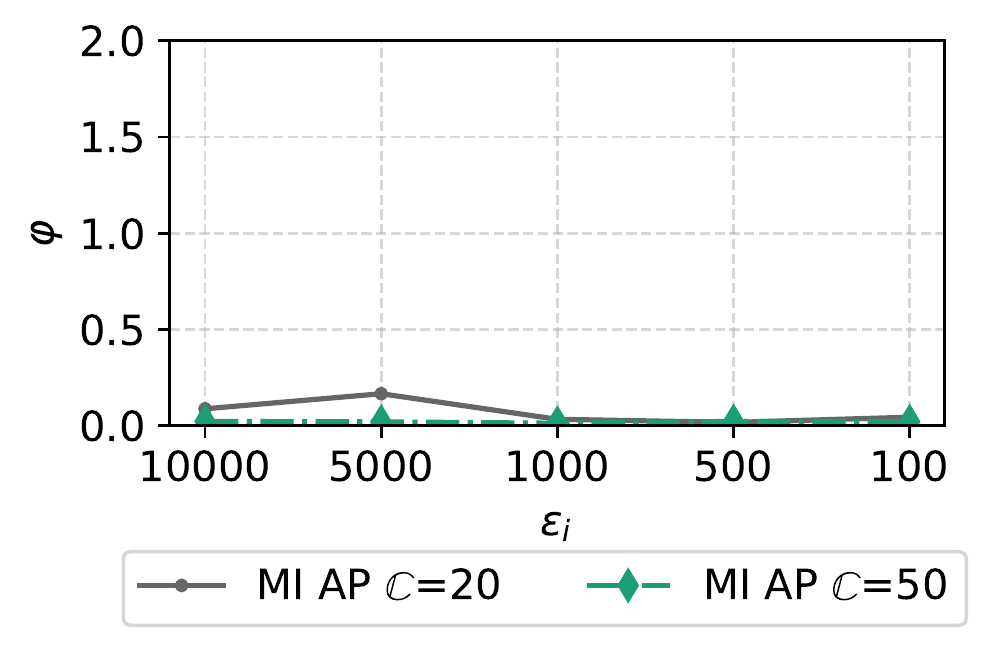}
        	\subcaption{LFW LDP.}
        	\label{fig:discussion:phi:ldpfull:lfw}
        \end{subfigure}
        \\
    	\begin{subfigure}{0.33\linewidth}
        	\includegraphics[width=\linewidth]{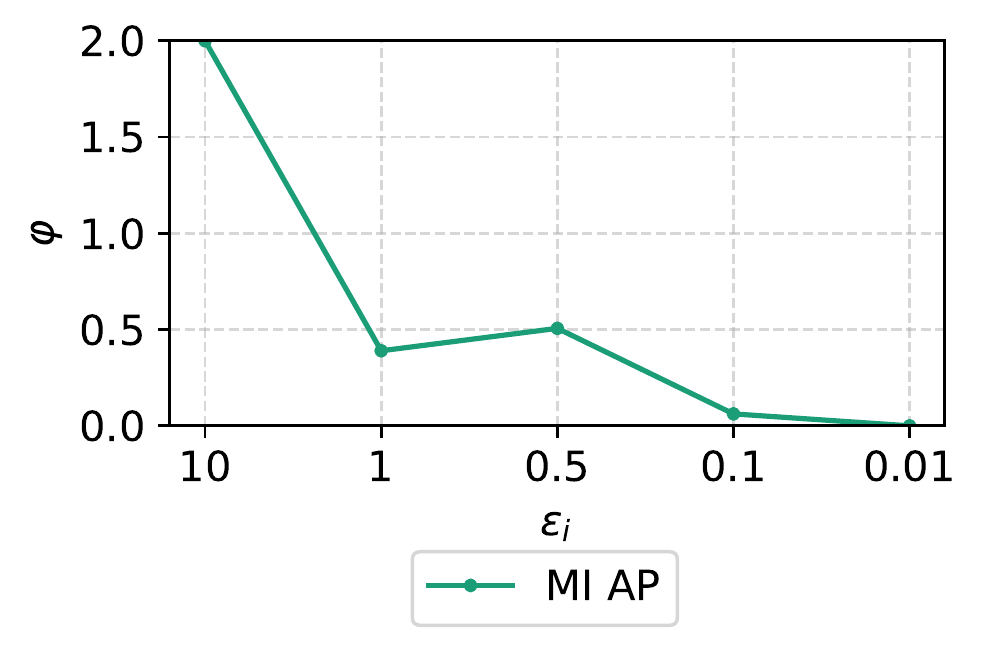}
    	    \subcaption{MS LDP.}
    	    \label{fig:discussion:phi:ldpfull:ms}
    	\end{subfigure}%
		\begin{subfigure}{0.33\linewidth}
            \includegraphics[width=\linewidth]{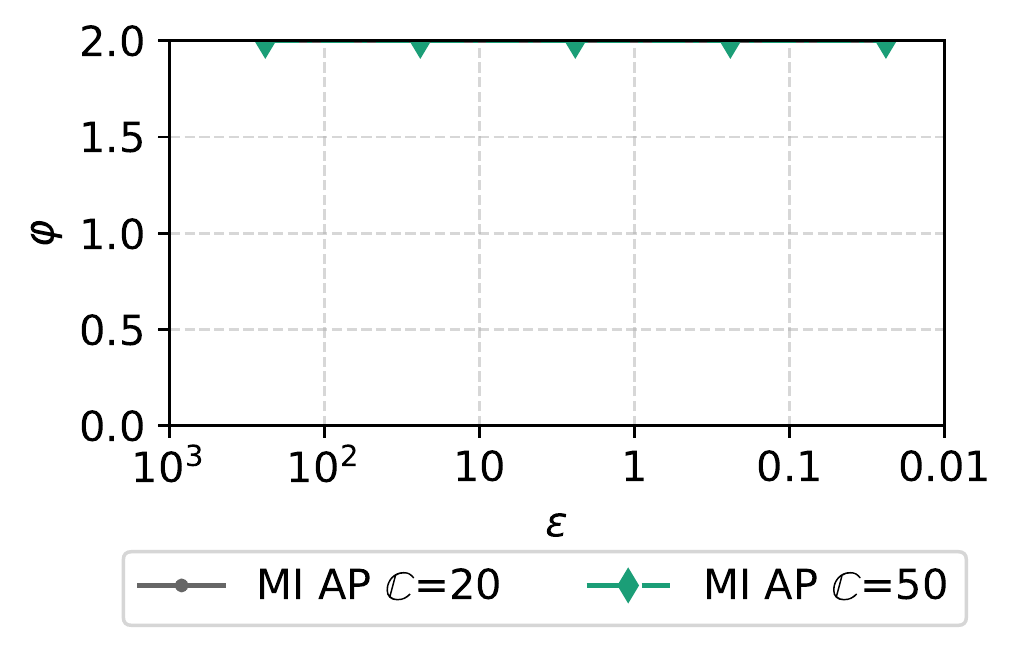}
        	\subcaption{LFW VAE-LDP.}
        	\label{fig:discussion:phi:vaeldpfull:lfw}
        \end{subfigure}%
    	\begin{subfigure}{0.33\linewidth}
        	\includegraphics[width=\linewidth]{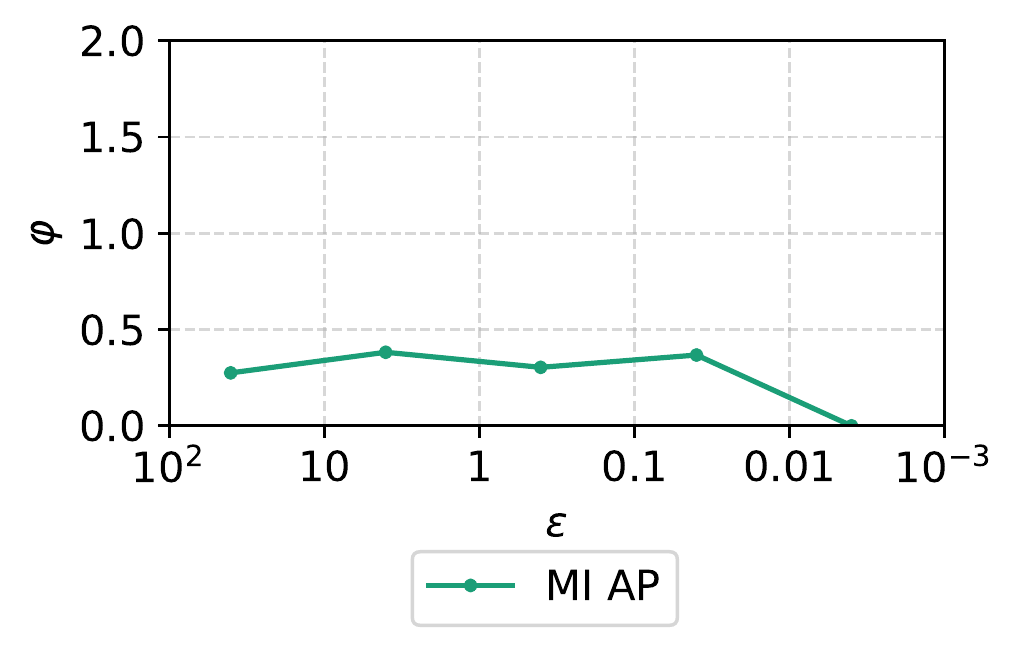}
    	    \subcaption{MS VAE-LDP.}
    	    \label{fig:discussion:phi:vaeldpfull:ms}
    	\end{subfigure}
	\caption{$\varphi$ for CDP, LDP and VAE-LDP, for LFW and MotionSense.}
	\label{fig:discussion:phi}
\end{figure}

\paragraph{Image data yields higher MI attack performance than time-series data.}
The reconstruction MI attack has been shown effective for image data in prior work~\cite{CYZF20,HHB19}, despite being fairly simple and only taking one metric for disparate behaviour of the target model into consideration. This is in line with the identified gap in image reconstruction for LFW and were able to exploit the gap by using SSIM as a distance measure for reconstruction MI attack. For MS we were not able to identify measure that provides equal success. Since activity measurements exhibit many ambiguities the target model learns to reconstruct relative trends instead of concrete measurements that represent a specific movement. Therefore, the target model generalizes more and is less prone to MI attacks. Additionally, previous research~\cite{NSH19,SSSS17} has shown that large datasets with few classes are generally less vulnerable to MI attacks.
\paragraph{Small noise yields favorable relative privacy-accuracy trade-off for image data.}
For CDP and image data we recommend using as little noise as possible. The relative accuracy drop for \cali{DS} largely exceeds the performance loss for $\Ami$ throughout the CDP experiments for LFW. This trend is illustrated in Figure~\ref{fig:discussion:phi:cdp:lfw} which highlights that the drop in target classifier test accuracy is always larger than the privacy gain by reduced MI AP. For MS the reconstruction MI attack only achieves a performance close to random guessing already against original data. Hence, small DP noise is already sufficient to push the MI AP to random guessing. This is reflected in Figure~\ref{fig:discussion:phi:cdp:ms}, where we see an optimal $\varphi$ already for $z=0.001$. Similarly for LDP Figures~\ref{fig:discussion:phi:ldpfull:lfw} and~\ref{fig:discussion:phi:ldpfull:ms} show only few favorable $\varphi$ for both datasets and settings. These few favorable trade-offs again indicate that differentially private image pixelization and the Laplace mechanism disproportionately harm model accuracy over protecting privacy. Compared to CDP, LDP shows better trade-offs for small privacy parameter. However, \cali{DS} generally gives up more accuracy compared to the gain in privacy.
\paragraph{VAE-LDP outperforms LDP and CDP w.r.t.~the relative privacy-accuracy trade-off.}
In our experiments, the VAE-LDP yielded the best trade-off between target classifier test accuracy and MI AP.
This finding is supported by $\varphi$ depicted in Figures~\ref{fig:discussion:phi:vaeldpfull:lfw} and ~\ref{fig:discussion:phi:vaeldpfull:ms}. We identified the interaction between the perturbation models, that retain essential image features, and the targeted classification task as primary reason for the superior trade-off. $\varphi$ for the VAE-LDP experiments highlight that small noise bounds are protecting from the reconstruction MI attack. For larger noise bounds however $\varphi$ only offers limited informative value since the MI AP pivots around random guessing while the target classifier test accuracy is bound by the overall classification baseline.
\paragraph{VAE are highly susceptible to noise introduced during training.}
Our results indicate that CDP leads to a regularization effect and directly addresses a key driver for MI AP. However, CDP also required additional hyperparameter optimization and increases computational cost. LDP mechanisms consume information within the data to foster protection and hence the test accuracy decrease heavily depends on how the LDP mechanism alters the training data. For example, differentially private image pixelization damages the structures of images to preserve privacy. The more information is consumed by the LDP mechanism, the worse the target classifier test accuracy becomes. This effect is clearly visible for the MS dataset, where the decrease in target classifier accuracy is similar to the overall classification baseline. When this characteristic is present MI is affected mostly as a consequence of diminishing model performance. This is facilitated by the lack of regularization effect which keeps a present relative gap for the MI attacks to exploit. The VAE-LDP mechanism preserves essential features of the LFW dataset during perturbation. The preservation of essential features are beneficial to the overall classification task as the test accuracy remains high while the MI AP decreases.
\section{Related Work}
\label{sec:rel_work}
We discuss related work from three categories. First, we briefly discuss generative models and accuracy metrics for generative models. Second, we provide background on differential privacy in generative models. Third, we introduce related work on membership inference attacks against generative models.

Generative Adversarial Networks by Goodfellow et al.~\cite{GPM+14} represent an alternative to VAE. We focus on VAE since VAE in comparison to GAN were observed to be more prone to MI attacks~\cite{HHB19}. Salimans et al.~\cite{SGZ+16} introduce Inception Score to automatically evaluate the utility of sampled images from generative models. The main advantage of Inception Score over other metrics such as SSIM is the correlation with human judgements. However, Barrat et al.~\cite{BS18} point out that Inception Score is foremost meaningful for the ImageNet dataset due to pre-training. Therefore, we consider the test accuracy of a target classifier to evaluate the VAE accuracy.

Torkzadehmahani et al.~\cite{TKP19} propose the DP-cGAN framework to generate differentially private data and labels. Similar to our work they train target classifiers on the generated data to evaluate model accuracy. We consider VAE with LDP and CDP. Jordon et al.~\cite{JYS19} extend the differentially private federated learning architecture PATE~\cite{PSM+18} to GAN. Similar to us, they analyze the accuracy of a target classifier for various privacy parameters, yet Jordon et al.~do not discuss privacy aside from privacy parameter \eps. Frigerio et al.~\cite{FOGD19} evaluate a CDP GAN for time series data also w.r.t.~MI attacks. We also consider LDP and quantify the trade-off between privacy and accuracy. Takahashi~\cite{TTOK20} propose an enhanced version of the DP-SGD for VAE by adjusting the noise that is injected to the loss terms. We use DP-Adam where their improvement is not applicable.

Hayes et al.~\cite{HMDC19} propose the LOGAN framework for MI attacks against GAN under various assumptions for the knowledge of $\Ami$. For their black-box attacks they train a separate discriminator model to distinguish between members and non-members. In contrast, we consider statistical MI attack models, allowing for MI attacks against generative models without the need to train a separate attack model. Hilprecht et al.~\cite{HHB19} propose Monte-Carlo MI attacks against GAN and VAE. We use their reconstruction MI attack and are the first to consider this attack under differential privacy. Chen et al.~\cite{CYZF20} extend the reconstruction MI attack to a partial black-box setting where $\Ami$ solely has access to the latent space $z$ but not the internal parameters of the generative model. Their attack composes different losses targeting various aspects of a model and takes the reconstruction as well as the latent representation into consideration. We ran all experiments within this paper also for their attack and the consideration of latent representation did lead to strictly weaker MI AP. The gradient matching attack of Zhu et al.~\cite{ZLH19} strives for reconstruction of training data from publicly available gradients. In contrast, we focus on the identification of training data. 
\section{Conclusion}
\label{sec:conc}
We evaluated a validation framework for quantifying the relative privacy-accuracy trade-off for VAE. We used the framework to compare two LDP and one CDP mechanism for image and time series data w.r.t.~their privacy-accuracy trade-off. In particular the LFW image recognition dataset was very susceptible to the reconstruction MI attack whereas the MotionSense activity recognition dataset with more records and less classes was mostly resistant to MI. The CDP mechanism offered a more consistent decrease of MI attack performance whereas the LDP mechanisms showed varying levels of protection depending on chosen privacy parameter and setting. The relative privacy-accuracy trade-off highlights that protection often comes at a disproportionately high accuracy cost.
\newpage
\section*{Appendix}
\label{sec:append}
\begin{table}[htb]
    \caption{Target Classifier hyperparameters.}
    \label{tab:experiments:target_classifier_hyper}
    \centering
    \begin{adjustbox}{width=1\textwidth}
\begin{tabular}{|ccc|ccccc|ccccc|}
\hline
\multicolumn{2}{|c|}{\multirow{2}{*}{}} & \multirow{2}{*}{\begin{tabular}[c]{@{}c@{}}Orig.,\\ CDP\end{tabular}} & \multicolumn{5}{c|}{LDP} & \multicolumn{5}{c|}{VAE-LDP} \\ \cline{4-13} 
\multicolumn{2}{|c|}{} &  & \multicolumn{1}{c|}{10000} & \multicolumn{1}{c|}{5000} & \multicolumn{1}{c|}{1000} & \multicolumn{1}{c|}{500} & 100 & \multicolumn{1}{c|}{0.1} & \multicolumn{1}{c|}{1} & \multicolumn{1}{c|}{10} & \multicolumn{1}{c|}{100} & 1000 \\ \hline
\multicolumn{1}{|c|}{\multirow{4}{*}{LFW20}} & \multicolumn{1}{c|}{learning rate} & 2.4e-05 & \multicolumn{1}{c|}{2.44e-4} & \multicolumn{1}{c|}{8.58e-05} & \multicolumn{1}{c|}{3.66e-05} & \multicolumn{1}{c|}{2.35e-4} & 1.43e-05 & \multicolumn{1}{c|}{4.03e-4} & \multicolumn{1}{c|}{1.42e-4} & \multicolumn{1}{c|}{9.34e-05} & \multicolumn{1}{c|}{1.39e-4} & 1.38e-3 \\ \cline{2-13} 
\multicolumn{1}{|c|}{} & \multicolumn{1}{c|}{batch size} & 16 & \multicolumn{1}{c|}{16} & \multicolumn{1}{c|}{16} & \multicolumn{1}{c|}{16} & \multicolumn{1}{c|}{16} & 64 & \multicolumn{1}{c|}{16} & \multicolumn{1}{c|}{64} & \multicolumn{1}{c|}{64} & \multicolumn{1}{c|}{16} & 64 \\ \cline{2-13} 
\multicolumn{1}{|c|}{} & \multicolumn{1}{c|}{epochs} & 33 & \multicolumn{1}{c|}{100} & \multicolumn{1}{c|}{10} & \multicolumn{1}{c|}{97} & \multicolumn{1}{c|}{16} & 24 & \multicolumn{1}{c|}{49} & \multicolumn{1}{c|}{34} & \multicolumn{1}{c|}{50} & \multicolumn{1}{c|}{45} & 46 \\ \cline{2-13} 
\multicolumn{1}{|c|}{} & \multicolumn{1}{c|}{test accuracy} & 0.98 & \multicolumn{1}{c|}{0.97} & \multicolumn{1}{c|}{0.97} & \multicolumn{1}{c|}{0.82} & \multicolumn{1}{c|}{0.55} & 0.28 & \multicolumn{1}{c|}{0.94} & \multicolumn{1}{c|}{0.93} & \multicolumn{1}{c|}{0.98} & \multicolumn{1}{c|}{1} & 1 \\ \hline
\multicolumn{1}{|c|}{\multirow{4}{*}{LFW50}} & \multicolumn{1}{c|}{learning rate} & 1e-05 & \multicolumn{1}{c|}{3.5e-05} & \multicolumn{1}{c|}{4.41e-4} & \multicolumn{1}{c|}{1.85e-4} & \multicolumn{1}{c|}{1.72e-4} & 1e-05 & \multicolumn{1}{c|}{9.24e-05} & \multicolumn{1}{c|}{3.29e-05} & \multicolumn{1}{c|}{7.39e-05} & \multicolumn{1}{c|}{9.76e-4} & 1.27e-4 \\ \cline{2-13} 
\multicolumn{1}{|c|}{} & \multicolumn{1}{c|}{batch size} & 16 & \multicolumn{1}{c|}{16} & \multicolumn{1}{c|}{64} & \multicolumn{1}{c|}{64} & \multicolumn{1}{c|}{64} & 64 & \multicolumn{1}{c|}{16} & \multicolumn{1}{c|}{16} & \multicolumn{1}{c|}{16} & \multicolumn{1}{c|}{64} & 32 \\ \cline{2-13} 
\multicolumn{1}{|c|}{} & \multicolumn{1}{c|}{epochs} & 100 & \multicolumn{1}{c|}{96} & \multicolumn{1}{c|}{100} & \multicolumn{1}{c|}{35} & \multicolumn{1}{c|}{90} & 21 & \multicolumn{1}{c|}{49} & \multicolumn{1}{c|}{37} & \multicolumn{1}{c|}{10} & \multicolumn{1}{c|}{32} & 20 \\ \cline{2-13} 
\multicolumn{1}{|c|}{} & \multicolumn{1}{c|}{test accuracy} & 0.95 & \multicolumn{1}{c|}{0.94} & \multicolumn{1}{c|}{0.93} & \multicolumn{1}{c|}{0.7} & \multicolumn{1}{c|}{0.41} & 0.2 & \multicolumn{1}{c|}{0.9} & \multicolumn{1}{c|}{0.91} & \multicolumn{1}{c|}{0.97} & \multicolumn{1}{c|}{1} & 1 \\ \hline
\multicolumn{1}{|c|}{\multirow{5}{*}{MS}} & \multicolumn{2}{c|}{$\eps_i$} & \multicolumn{1}{c|}{10} & \multicolumn{1}{c|}{1} & \multicolumn{1}{c|}{0.5} & \multicolumn{1}{c|}{0.1} & 0.01 & \multicolumn{1}{c|}{0.1} & \multicolumn{1}{c|}{1} & \multicolumn{1}{c|}{10} & \multicolumn{1}{c|}{100} & 1000 \\ \cline{2-13} 
\multicolumn{1}{|c|}{} & \multicolumn{1}{c|}{learning rate} & 9.8e-4 & \multicolumn{1}{c|}{9.37e-4} & \multicolumn{1}{c|}{7.26e-4} & \multicolumn{1}{c|}{7.72e-4} & \multicolumn{1}{c|}{9.87e-05} & 1.08e-05 & \multicolumn{1}{c|}{1.09e-3} & \multicolumn{1}{c|}{6.75e-4} & \multicolumn{1}{c|}{1.14e-4} & \multicolumn{1}{c|}{2.48e-3} & 3.71e-05 \\ \cline{2-13} 
\multicolumn{1}{|c|}{} & \multicolumn{1}{c|}{batch size} & 64 & \multicolumn{1}{c|}{64} & \multicolumn{1}{c|}{64} & \multicolumn{1}{c|}{16} & \multicolumn{1}{c|}{16} & 16 & \multicolumn{1}{c|}{256} & \multicolumn{1}{c|}{128} & \multicolumn{1}{c|}{32} & \multicolumn{1}{c|}{32} & 64 \\ \cline{2-13} 
\multicolumn{1}{|c|}{} & \multicolumn{1}{c|}{epochs} & 25 & \multicolumn{1}{c|}{25} & \multicolumn{1}{c|}{25} & \multicolumn{1}{c|}{25} & \multicolumn{1}{c|}{25} & 6 & \multicolumn{1}{c|}{21} & \multicolumn{1}{c|}{9} & \multicolumn{1}{c|}{23} & \multicolumn{1}{c|}{16} & 24 \\ \cline{2-13} 
\multicolumn{1}{|c|}{} & \multicolumn{1}{c|}{test accuracy} & 0.99 & \multicolumn{1}{c|}{0.98} & \multicolumn{1}{c|}{0.93} & \multicolumn{1}{c|}{0.8} & \multicolumn{1}{c|}{0.29} & 0.25 & \multicolumn{1}{c|}{0.68} & \multicolumn{1}{c|}{0.53} & \multicolumn{1}{c|}{0.39} & \multicolumn{1}{c|}{0.3} & 0.24 \\ \hline
\end{tabular}
    \end{adjustbox}
\end{table}
\begin{figure}[ht!]
	\centering
		\begin{subfigure}{0.4\linewidth}
            \includegraphics[width=\linewidth]{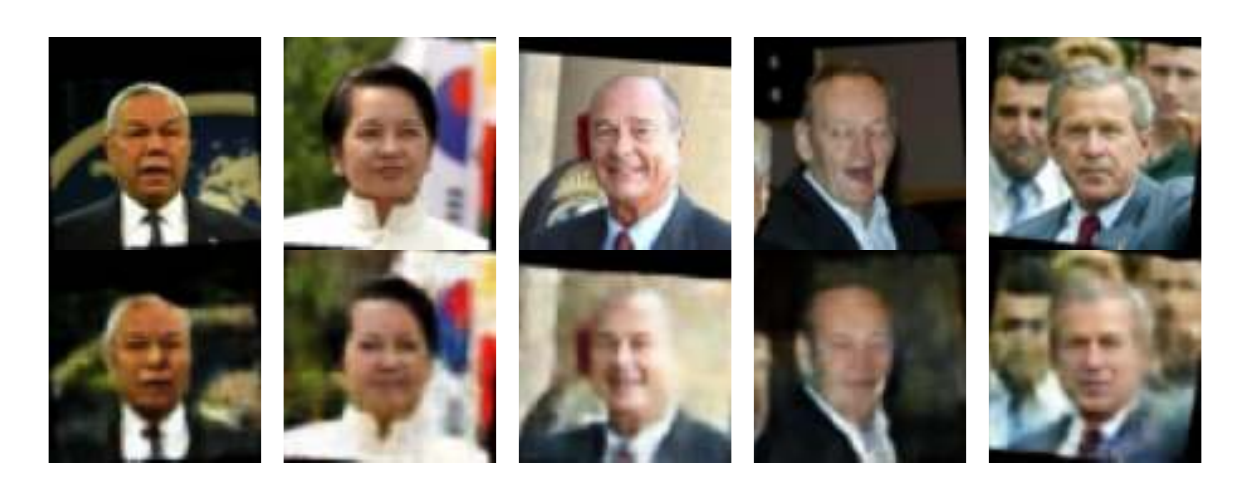}
        	\subcaption{Reconstructed training records.}
        	\label{fig:experiments:baseline:lfw:visual:train}
        \end{subfigure}
    	\begin{subfigure}{0.4\linewidth}
    	\includegraphics[width=\linewidth]{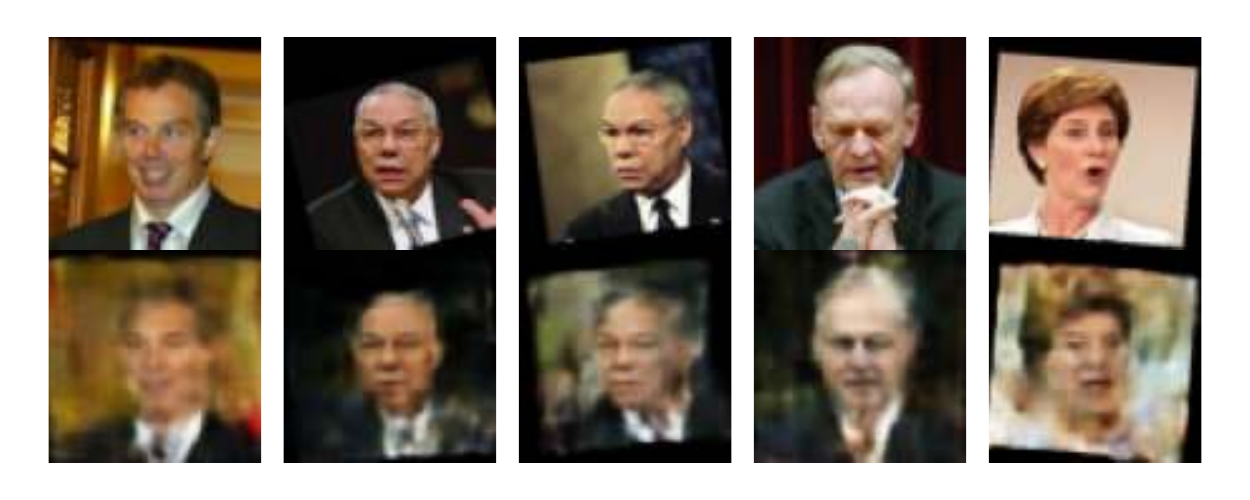}
    	\subcaption{Reconstructed test records.}
    	    \label{fig:experiments:baseline:lfw:visual:test}
    	\end{subfigure}
    	\\
		\begin{subfigure}{1\linewidth}
		\centering
            \includegraphics[width=0.8\linewidth]{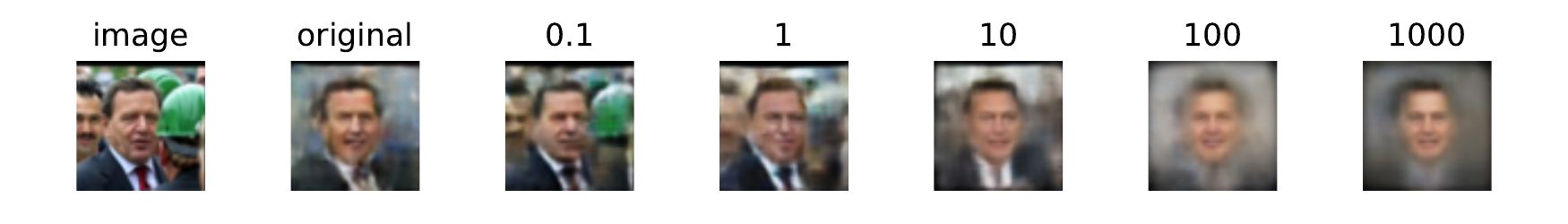}
        	\subcaption{VAE-LDP generated samples for LFW20.}
        	\label{fig:experiments:baseline:vaegenrecon:lfw20}
        \end{subfigure}
	\caption{Comparison of reconstructed records and generated samples.}
	\label{fig:experiments:samples}
\end{figure}
\begin{figure}[h!tb]
	\centering
        \includegraphics[width=0.45\linewidth]{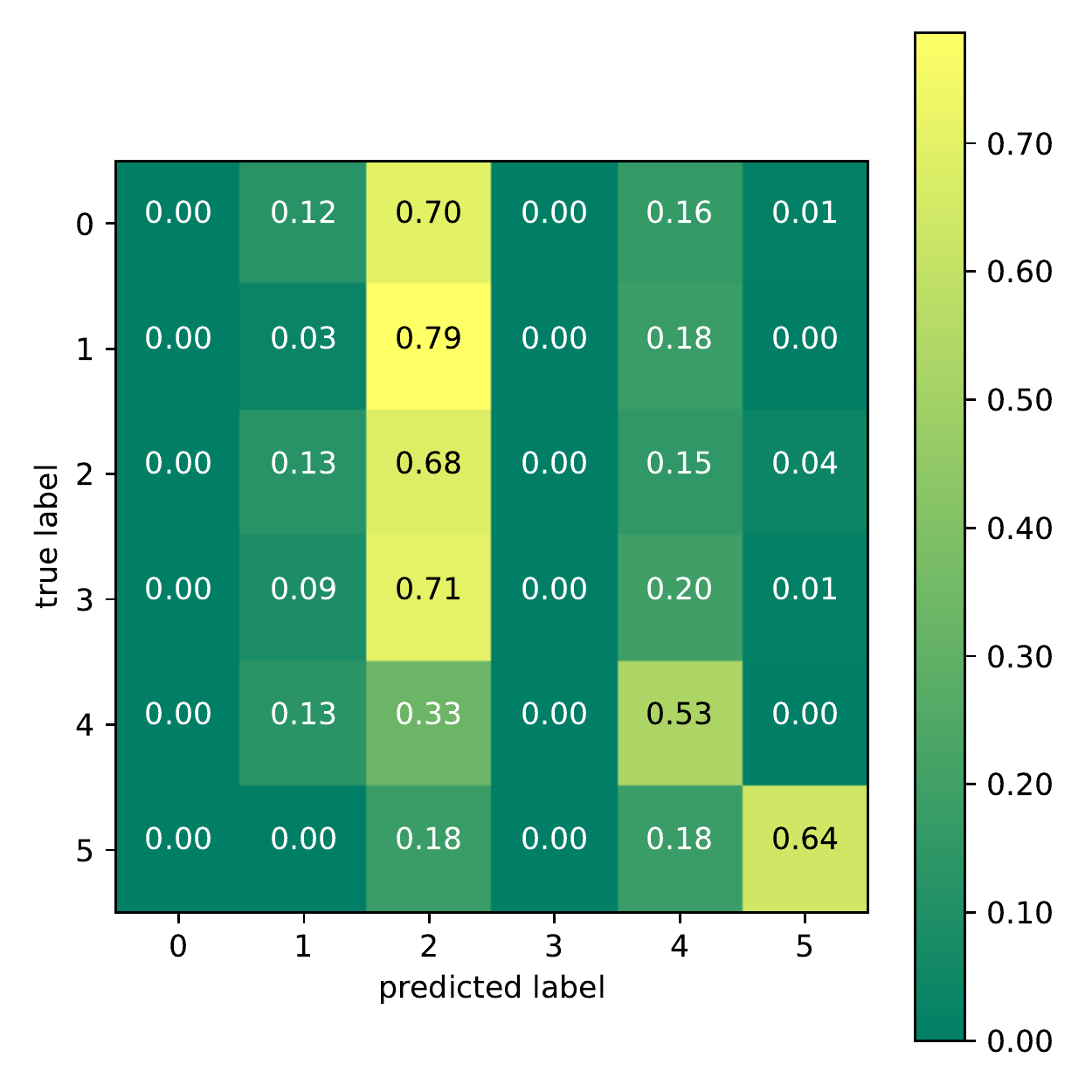}
	\caption{Confusion matrix for the target classifier for MotionSense CDP $z=1$.} 
	\label{fig:experiments:cdp:ms:cm}
\end{figure}
\begin{table}[h!tp]
    \caption{Target model hyperparameters, CDP and VAE-LDP $\epsdlt$, and LDP $\eps$.}
    \label{tab:experiments:target_model}
    \centering
    \begin{adjustbox}{width=1\textwidth}
\begin{tabular}{|cccccccc|}
\hline
\multicolumn{2}{|c|}{} & \multicolumn{1}{c|}{Orig.} & \multicolumn{5}{c|}{CDP} \\ \hline
\multicolumn{2}{|c|}{$z$} & \multicolumn{1}{c|}{} & \multicolumn{1}{c|}{0.001} & \multicolumn{1}{c|}{0.01} & \multicolumn{1}{c|}{0.1} & \multicolumn{1}{c|}{0.5} & 1 \\ \hline
\multicolumn{1}{|c|}{\multirow{6}{*}{LFW20}} & \multicolumn{1}{c|}{learning rate} & \multicolumn{1}{c|}{5.57e-4} & \multicolumn{1}{c|}{4.67e-4} & \multicolumn{1}{c|}{1.82e-4} & \multicolumn{1}{c|}{1.1e-4} & \multicolumn{1}{c|}{5.62e-4} & 4.01e-05 \\ \cline{2-8} 
\multicolumn{1}{|c|}{} & \multicolumn{1}{c|}{batch size} & \multicolumn{1}{c|}{16} & \multicolumn{1}{c|}{32} & \multicolumn{1}{c|}{16} & \multicolumn{1}{c|}{64} & \multicolumn{1}{c|}{16} & 32 \\ \cline{2-8} 
\multicolumn{1}{|c|}{} & \multicolumn{1}{c|}{epochs} & \multicolumn{1}{c|}{1000} & \multicolumn{5}{c|}{1000} \\ \cline{2-8} 
\multicolumn{1}{|c|}{} & \multicolumn{1}{c|}{$\cali{C}$, microbatch} & \multicolumn{1}{c|}{-} & \multicolumn{5}{c|}{0.03, 4} \\ \cline{2-8} 
\multicolumn{1}{|c|}{} & \multicolumn{1}{c|}{$\epsdlt$} & \multicolumn{1}{c|}{-} & \multicolumn{1}{c|}{\begin{tabular}[c]{@{}c@{}}(15948925900.96, \\ 1.08e-03)\end{tabular}} & \multicolumn{1}{c|}{\begin{tabular}[c]{@{}c@{}}(321853746.70, \\ 1.08e-03)\end{tabular}} & \multicolumn{1}{c|}{\begin{tabular}[c]{@{}c@{}}(372950.22, \\ 1.08e-03)\end{tabular}} & \multicolumn{1}{c|}{\begin{tabular}[c]{@{}c@{}}(295.07, \\ 1.08e-03)\end{tabular}} & \begin{tabular}[c]{@{}c@{}}(56.41, \\ 1.08e-03)\end{tabular} \\ \cline{2-8} 
\multicolumn{1}{|c|}{} & \multicolumn{1}{c|}{train-test gap} & \multicolumn{1}{c|}{260.3} & \multicolumn{1}{c|}{210.6} & \multicolumn{1}{c|}{62.5} & \multicolumn{1}{c|}{13.8} & \multicolumn{1}{c|}{17.9} & 10.5 \\ \hline
\multicolumn{1}{|c|}{\multirow{6}{*}{LFW50}} & \multicolumn{1}{c|}{learning rate} & \multicolumn{1}{c|}{5.88e-4} & \multicolumn{1}{c|}{2.15e-4} & \multicolumn{1}{c|}{1.04e-4} & \multicolumn{1}{c|}{5.14e-05} & \multicolumn{1}{c|}{1.93e-4} & 6.86e-4 \\ \cline{2-8} 
\multicolumn{1}{|c|}{} & \multicolumn{1}{c|}{batch size} & \multicolumn{1}{c|}{16} & \multicolumn{2}{c|}{16} & \multicolumn{3}{c|}{32} \\ \cline{2-8} 
\multicolumn{1}{|c|}{} & \multicolumn{1}{c|}{epochs} & \multicolumn{1}{c|}{1000} & \multicolumn{5}{c|}{1000} \\ \cline{2-8} 
\multicolumn{1}{|c|}{} & \multicolumn{1}{c|}{$\cali{C}$, microbatch} & \multicolumn{1}{c|}{-} & \multicolumn{5}{c|}{0.02, 4} \\ \cline{2-8} 
\multicolumn{1}{|c|}{} & \multicolumn{1}{c|}{$(\epsilon,\delta)$} & \multicolumn{1}{c|}{-} & \multicolumn{1}{c|}{\begin{tabular}[c]{@{}c@{}}(47295786259.73, \\ 7.27e-04)\end{tabular}} & \multicolumn{1}{c|}{\begin{tabular}[c]{@{}c@{}}(468786259.73, \\ 7.27e-04)\end{tabular}} & \multicolumn{1}{c|}{\begin{tabular}[c]{@{}c@{}}(681781.38, \\ 7.27e-04)\end{tabular}} & \multicolumn{1}{c|}{\begin{tabular}[c]{@{}c@{}}(353.74, \\ 7.27e-04)\end{tabular}} & \begin{tabular}[c]{@{}c@{}}(43.31, \\ 7.27e-04)\end{tabular} \\ \cline{2-8} 
\multicolumn{1}{|c|}{} & \multicolumn{1}{c|}{train-test gap} & \multicolumn{1}{c|}{259.4} & \multicolumn{1}{c|}{195.4} & \multicolumn{1}{c|}{29.4} & \multicolumn{1}{c|}{9.2} & \multicolumn{1}{c|}{7.1} & 33 \\ \hline
\multicolumn{1}{|c|}{\multirow{6}{*}{MS}} & \multicolumn{1}{c|}{learning rate} & \multicolumn{6}{c|}{1e-3} \\ \cline{2-8} 
\multicolumn{1}{|c|}{} & \multicolumn{1}{c|}{batch size} & \multicolumn{6}{c|}{32} \\ \cline{2-8} 
\multicolumn{1}{|c|}{} & \multicolumn{1}{c|}{epochs} & \multicolumn{6}{c|}{1000} \\ \cline{2-8} 
\multicolumn{1}{|c|}{} & \multicolumn{1}{c|}{$\cali{C}$, microbatch} & \multicolumn{1}{c|}{-} & \multicolumn{5}{c|}{3.4e-5, 4} \\ \cline{2-8} 
\multicolumn{1}{|c|}{} & \multicolumn{1}{c|}{$\epsdlt$} & \multicolumn{1}{c|}{-} & \multicolumn{1}{c|}{\begin{tabular}[c]{@{}c@{}}(120986947509.93, \\ 1.42e-04)\end{tabular}} & \multicolumn{1}{c|}{\begin{tabular}[c]{@{}c@{}}(1196947509.93, \\ 1.42e-04)\end{tabular}} & \multicolumn{1}{c|}{\begin{tabular}[c]{@{}c@{}}(1093201.38, \\ 1.42e-04)\end{tabular}} & \multicolumn{1}{c|}{\begin{tabular}[c]{@{}c@{}}(137.57, \\ 1.42e-04)\end{tabular}} & \begin{tabular}[c]{@{}c@{}}(15.73, \\ 1.42e-04)\end{tabular} \\ \cline{2-8} 
\multicolumn{1}{|c|}{} & \multicolumn{1}{c|}{train-test gap} & \multicolumn{1}{c|}{0.7} & \multicolumn{1}{c|}{0.4} & \multicolumn{1}{c|}{0.3} & \multicolumn{1}{c|}{0.1} & \multicolumn{1}{c|}{0.1} & 0 \\ \hline
\multicolumn{8}{|c|}{LDP} \\ \hline
\multicolumn{3}{|c|}{$\epsilon_i$} & \multicolumn{1}{c|}{10000} & \multicolumn{1}{c|}{5000} & \multicolumn{1}{c|}{1000} & \multicolumn{1}{c|}{500} & 100 \\ \hline
\multicolumn{1}{|c|}{\multirow{5}{*}{LFW20}} & \multicolumn{2}{c|}{learning rate} & \multicolumn{1}{c|}{9.22e-4} & \multicolumn{1}{c|}{1.52e-4} & \multicolumn{1}{c|}{2.13e-4} & \multicolumn{1}{c|}{1.14e-4} & 1e-3 \\ \cline{2-8} 
\multicolumn{1}{|c|}{} & \multicolumn{2}{c|}{batch size} & \multicolumn{1}{c|}{32} & \multicolumn{1}{c|}{16} & \multicolumn{1}{c|}{64} & \multicolumn{1}{c|}{32} & 16 \\ \cline{2-8} 
\multicolumn{1}{|c|}{} & \multicolumn{2}{c|}{epochs} & \multicolumn{5}{c|}{1000} \\ \cline{2-8} 
\multicolumn{1}{|c|}{} & \multicolumn{2}{c|}{$\epsilon$} & \multicolumn{1}{c|}{5.718e+07} & \multicolumn{1}{c|}{2.859e+07} & \multicolumn{1}{c|}{5.718e+06} & \multicolumn{1}{c|}{2.859e+06} & 571800 \\ \cline{2-8} 
\multicolumn{1}{|c|}{} & \multicolumn{2}{c|}{train-test gap} & \multicolumn{1}{c|}{267} & \multicolumn{1}{c|}{265} & \multicolumn{1}{c|}{224} & \multicolumn{1}{c|}{160} & 123 \\ \hline
\multicolumn{1}{|c|}{\multirow{5}{*}{LFW50}} & \multicolumn{2}{c|}{learning rate} & \multicolumn{1}{c|}{4.61e-4} & \multicolumn{1}{c|}{2.41e-4} & \multicolumn{1}{c|}{4.31e-4} & \multicolumn{1}{c|}{1.19e-05} & 1e-05 \\ \cline{2-8} 
\multicolumn{1}{|c|}{} & \multicolumn{2}{c|}{batch size} & \multicolumn{2}{c|}{16} & \multicolumn{3}{c|}{64} \\ \cline{2-8} 
\multicolumn{1}{|c|}{} & \multicolumn{2}{c|}{epochs} & \multicolumn{5}{c|}{1000} \\ \cline{2-8} 
\multicolumn{1}{|c|}{} & \multicolumn{2}{c|}{$\epsilon$} & \multicolumn{1}{c|}{8.319e+07} & \multicolumn{1}{c|}{4.1595e+07} & \multicolumn{1}{c|}{8.319e+06} & \multicolumn{1}{c|}{4.1595e+06} & 831900 \\ \cline{2-8} 
\multicolumn{1}{|c|}{} & \multicolumn{2}{c|}{train-test gap} & \multicolumn{1}{c|}{272} & \multicolumn{1}{c|}{264} & \multicolumn{1}{c|}{204} & \multicolumn{1}{c|}{21} & 5 \\ \hline
\multicolumn{3}{|c|}{$\epsilon_i$} & \multicolumn{1}{c|}{10} & \multicolumn{1}{c|}{1} & \multicolumn{1}{c|}{0.5} & \multicolumn{1}{c|}{0.1} & 0.01 \\ \hline
\multicolumn{1}{|c|}{\multirow{5}{*}{MS}} & \multicolumn{2}{c|}{learning rate} & \multicolumn{5}{c|}{1e-3} \\ \cline{2-8} 
\multicolumn{1}{|c|}{} & \multicolumn{2}{c|}{batch size} & \multicolumn{5}{c|}{32} \\ \cline{2-8} 
\multicolumn{1}{|c|}{} & \multicolumn{2}{c|}{epochs} & \multicolumn{5}{c|}{1000} \\ \cline{2-8} 
\multicolumn{1}{|c|}{} & \multicolumn{2}{c|}{$\epsilon$} & \multicolumn{1}{c|}{706190} & \multicolumn{1}{c|}{70619} & \multicolumn{1}{c|}{35309.5} & \multicolumn{1}{c|}{7061.9} & 706.19 \\ \cline{2-8} 
\multicolumn{1}{|c|}{} & \multicolumn{2}{c|}{train-test gap} & \multicolumn{1}{c|}{0.7} & \multicolumn{1}{c|}{0.9} & \multicolumn{1}{c|}{2.7} & \multicolumn{1}{c|}{4.4} & 4.8 \\ \hline
\multicolumn{8}{|c|}{VAE-LDP} \\ \hline
\multicolumn{3}{|c|}{$\sigma$} & \multicolumn{1}{c|}{0.1} & \multicolumn{1}{c|}{1} & \multicolumn{1}{c|}{10} & \multicolumn{1}{c|}{100} & 1000 \\ \hline
\multicolumn{1}{|c|}{\multirow{5}{*}{LFW20}} & \multicolumn{2}{c|}{learning rate} & \multicolumn{5}{c|}{5.57e-4} \\ \cline{2-8} 
\multicolumn{1}{|c|}{} & \multicolumn{2}{c|}{batch size} & \multicolumn{5}{c|}{16} \\ \cline{2-8} 
\multicolumn{1}{|c|}{} & \multicolumn{2}{c|}{epochs} & \multicolumn{5}{c|}{1000} \\ \cline{2-8} 
\multicolumn{1}{|c|}{} & \multicolumn{2}{c|}{$\epsdlt$} & \multicolumn{1}{c|}{\begin{tabular}[c]{@{}c@{}}(2366.15, \\ 5.25e-04)\end{tabular}} & \multicolumn{1}{c|}{\begin{tabular}[c]{@{}c@{}}(236.61, \\ 5.25e-04)\end{tabular}} & \multicolumn{1}{c|}{\begin{tabular}[c]{@{}c@{}}(23.66, \\ 5.25e-04)\end{tabular}} & \multicolumn{1}{c|}{\begin{tabular}[c]{@{}c@{}}(2.37, \\ 5.25e-04)\end{tabular}} & \begin{tabular}[c]{@{}c@{}}(0.24, \\ 5.25e-04)\end{tabular} \\ \cline{2-8} 
\multicolumn{1}{|c|}{} & \multicolumn{2}{c|}{train-test gap} & \multicolumn{1}{c|}{156} & \multicolumn{1}{c|}{145} & \multicolumn{1}{c|}{64} & \multicolumn{1}{c|}{3} & 2 \\ \hline
\multicolumn{1}{|c|}{\multirow{5}{*}{LFW50}} & \multicolumn{2}{c|}{learning rate} & \multicolumn{5}{c|}{5.88e-4} \\ \cline{2-8} 
\multicolumn{1}{|c|}{} & \multicolumn{2}{c|}{batch size} & \multicolumn{5}{c|}{16} \\ \cline{2-8} 
\multicolumn{1}{|c|}{} & \multicolumn{2}{c|}{epochs} & \multicolumn{5}{c|}{1000} \\ \cline{2-8} 
\multicolumn{1}{|c|}{} & \multicolumn{2}{c|}{$\epsdlt$} & \multicolumn{1}{c|}{\begin{tabular}[c]{@{}c@{}}(2422.52, \\ 3.61e-04)\end{tabular}} & \multicolumn{1}{c|}{\begin{tabular}[c]{@{}c@{}}(242.25, \\ 3.61e-04)\end{tabular}} & \multicolumn{1}{c|}{\begin{tabular}[c]{@{}c@{}}(24.23, \\ 3.61e-04)\end{tabular}} & \multicolumn{1}{c|}{\begin{tabular}[c]{@{}c@{}}(2.42, \\ 3.61e-04)\end{tabular}} & \begin{tabular}[c]{@{}c@{}}(0.24, \\ 3.61e-04)\end{tabular} \\ \cline{2-8} 
\multicolumn{1}{|c|}{} & \multicolumn{2}{c|}{train-test gap} & \multicolumn{1}{c|}{168} & \multicolumn{1}{c|}{158} & \multicolumn{1}{c|}{68} & \multicolumn{1}{c|}{4} & 3 \\ \hline
\multicolumn{1}{|c|}{\multirow{5}{*}{MS}} & \multicolumn{2}{c|}{learning rate} & \multicolumn{5}{c|}{1e-3} \\ \cline{2-8} 
\multicolumn{1}{|c|}{} & \multicolumn{2}{c|}{batch size} & \multicolumn{5}{c|}{32} \\ \cline{2-8} 
\multicolumn{1}{|c|}{} & \multicolumn{2}{c|}{epochs} & \multicolumn{5}{c|}{1000} \\ \cline{2-8} 
\multicolumn{1}{|c|}{} & \multicolumn{2}{c|}{$\epsdlt$} & \multicolumn{1}{c|}{\begin{tabular}[c]{@{}c@{}}(404.96, \\ 1.42e-05)\end{tabular}} & \multicolumn{1}{c|}{\begin{tabular}[c]{@{}c@{}}(40.50, \\ 1.42e-05)\end{tabular}} & \multicolumn{1}{c|}{\begin{tabular}[c]{@{}c@{}}(4.05, \\ 1.42e-05)\end{tabular}} & \multicolumn{1}{c|}{\begin{tabular}[c]{@{}c@{}}(0.40, \\ 1.42e-05)\end{tabular}} & \begin{tabular}[c]{@{}c@{}}(0.04, \\ 1.42e-05)\end{tabular} \\ \cline{2-8} 
\multicolumn{1}{|c|}{} & \multicolumn{2}{c|}{train-test gap} & \multicolumn{1}{c|}{0.1} & \multicolumn{1}{c|}{0} & \multicolumn{1}{c|}{0.2} & \multicolumn{1}{c|}{0.1} & 0 \\ \hline
\end{tabular}
    \end{adjustbox}
\end{table}

\bibliographystyle{abbrv}
\bibliography{arxiv.bib}

\end{document}